\documentclass[aps,prb,twocolumn, showpacs,longbibliography,notitlepage]{revtex4-1}

\usepackage{amsmath}
\usepackage{graphicx}
\usepackage{lmodern}
\usepackage{amsmath}
\usepackage{bm}
\usepackage{hyperref}\usepackage{url}
\usepackage{subfigure}%
\usepackage{dsfont}

\usepackage{amsbsy}
\usepackage{dcolumn}
\usepackage{amsthm}
\usepackage{bm}
\usepackage{esint}
\usepackage{multirow}
\usepackage{hyperref}
 
\usepackage{cleveref}

\usepackage{mathrsfs}
\usepackage{amsfonts}
\usepackage{amsbsy}
\usepackage{dcolumn}
\usepackage{bm}
\usepackage{multirow}
\usepackage{color}

\usepackage{color}
\usepackage{amssymb}
\newcommand{\beq} {\begin{equation}}
\newcommand{\eeq} {\end{equation}}
\newcommand{\bea} {\begin{eqnarray}}
\newcommand{\eea} {\end{eqnarray}}
\newcommand{\be} {\begin{equation}}
\newcommand{\ee} {\end{equation}}
\renewcommand{\(}{\left(}
\renewcommand{\)}{\right)}
\renewcommand{\[}{\left[}
\renewcommand{\]}{\right]}

\DeclareMathOperator{\sgn}{sgn}

\begin{document}

\title {Topological density-wave states in a particle-hole symmetric Weyl metal}
\author{Yuxuan Wang}
\author{Peng Ye}
\affiliation{Department of Physics and Institute for Condensed Matter Theory, University of Illinois at Urbana-Champaign, 1110 West Green Street, Urbana, Illinois 61801-3080, USA }

\begin{abstract}
We study the instabilities of a particle-hole symmetric Weyl metal with both electron and hole Fermi surfaces (FS) around the Weyl points.  For a repulsive interaction, we find that the leading instability is towards a longitudinal spin-density-wave (SDW$_z$) order. Besides, there exist three degenerate subleading instabilities: a charge-density-wave (CDW) instability and two transverse spin-density-wave (SDW$_{x,y}$) instabilities. For an attractive interaction the leading instabilities are towards two pair-density-wave orders (PDW) which pair the two FS's separately. Both the PDW and SDW$_z$ order parameters fully gap out the FS's, while the CDW and SDW$_{x,y}$ ones  leave line nodes on both FS's. For the SDW$_z$ and the PDW states, the surface Fermi arc in the metallic state evolves to a chiral Fermi line which passes the projection of the Weyl points and  traverses the full momentum space. For the CDW state, the line node projects to a ``drumhead" band localized on the surface, which can lead to a topological charge polarization. We verify the surface states by computing the angular-resolved photoemission spectroscopy data.
\end{abstract}
\date{\today}

\maketitle

\section{ Introduction}
\begin{figure}
 \includegraphics[width=0.81\columnwidth]{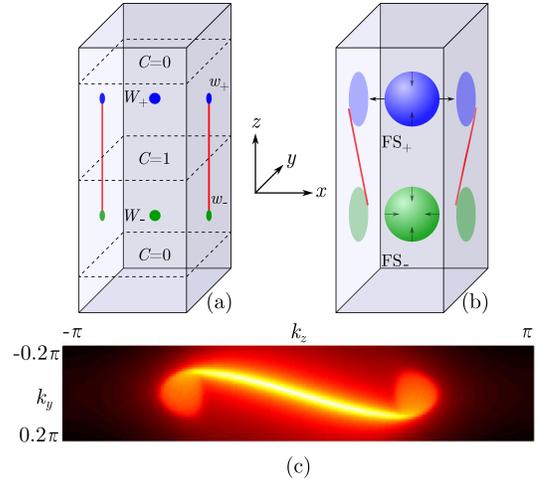}
 \caption{(Color online) (a) The bulk and surface Brillouin zones (BZ) of a WSM, with two Weyl {points} shown in blue ($W_{+}$) and green ($W_-$) {respectively}.  Given a $k_z$, {each} 2D {slice}  can be viewed as a Chern insulator with Chern number $C$. The surface band structure exhibits  Fermi arcs (denoted by red lines), each connecting the surface projections (denoted by $w_{\pm}$) of the bulk Weyl points. (b) The BZ of a $\mathcal{C}$-WM, with a hole FS (blue) and an electron FS (green), denoted as FS$_{\pm}$. Both FS$_\pm$ are spin-textured, as shown by the arrows. The surface Fermi arcs terminate on the surface projections of FS$_\pm$. (c) Simulation of the ARPES data on a $yz$-surface for a $\mathcal{C}$-WM with $Q=\pi/2$, showing the Fermi arc that terminates at the projections of the FS$_{\pm}$. For clarity the color coding is at logarithmic scale.} 
   \label{fig1}
 \end{figure}

The study of topological phases of matter 
has been one of the most remarkable achievements  in condensed matter physics over the {past decade}.  {Recently, this line of thinking has been extended from gapped systems to semimetallic systems}, such as  Weyl semimetals (WSM) \cite{xwan11,xugang11,burkov_balents2011,YLR2011,Xu613,Lv2015,Xu:2015aa,Lv:2015aa,Yang:2015aa,Xue1501092,Liu:2016aa,Belopolski2016,Weng:2016aa,Liang:2016aa,Huang:2015aa, FengNbP,Soluyanov:2015aa,type2a,type2b,type2c,type2d,type2e,bhyanPRB2015} and Dirac semimetals \cite{Wang_Dirac2012,Wang_Dirac2013,Liu864,cava2014,nem_dirac,jim2015}.   
{The bulk band structure of WSMs is characterized by 
Weyl nodes that result from the linear touching of two bands in three dimensions. The low energy theory of the WSM is dominated by relativistic Weyl fermions around the Weyl points. According to the Nielsen-Ninomiya theorem \cite{NIELSEN198120,NIELSEN1981173}, in any lattice models of the WSM, the Weyl points can only appear in pairs of opposite chiralities.}
A WSM can also be viewed as the stacking of two-dimensional (2D) Chern insulators with Chern number $C$ in the momentum ($\bm{k}$) space~\cite{YLR2011}.
 Since a Weyl point is a source or sink of  Berry fluxes,   $C$ changes by one as the $\bm k$-space slice goes through a Weyl point. 
WSMs exhibit chiral  anomaly in the form of the anomalous Hall effect and the chiral magnetic effect~\cite{aji2012,12a,Son_2012,Grushin_2012,burkov_theta,hosur2013,goswami2013}. Besides, on its surfaces there exist ``Fermi arcs"~\cite{xwan11,Belopolski2016} that terminate at  the surface projections of the Weyl {points} of opposite chiralities [see Fig.~\ref{fig1}(a)], which has been recently observed in transition metal pnictides~\cite{Lv2015,Lv:2015aa,Yang:2015aa,Xue1501092,Liu:2016aa,Belopolski2016,Weng:2016aa,Liang:2016aa,Huang:2015aa, Xu:2015aa,FengNbP}.

An interesting extension to  the WSM is the 
Weyl metal (WM), where  Weyl points  evolve into 2D Fermi surfaces (FS) upon a shift in chemical potential.
  Each FS encloses a Weyl point, and each FS is spin-textured. 
  At low temperatures, with a small attractive interaction these FS's are subject to a number of instabilities in the particle-particle channel. 
  One such instability is towards a uniform superconducting (SC) order~\cite{Bednik2015,LuTanaka2015,Lihaldane:2015aa,Wang:2016aa,Cho_weyl_2012}. On the other hand, fermions on a given FS can form intra-FS pairs, leading to a pair-density-wave (PDW) state~\cite{Cho_weyl_2012,Bednik2015,susy_weyl,gilbert2016,rodrigo2015,you2}, which is similar to Fulde-Ferrel-Larkin-Ovchinnikov (FFLO) state but without an external field. Such a state has also been proposed and searched for in cuprate superconductors and many microscopic models~\cite{fradkin,palee,agterberg,pdw1,pdw2,pdw_pepin,davis_last}. However, it is usually secondary to the uniform SC order~\cite{Bednik2015}, at least at weak coupling.
   On the other hand, in the presence of the SC order, the surface Fermi arc was found  to receive an interesting reconstruction due to the change of the bulk band structure~\cite{LuTanaka2015,Lihaldane:2015aa}.

 In this paper, we study the density-wave instabilities in a particle-hole symmetric Weyl metal ($\mathcal{C}$-WM) and analyze the reconstruction of electronic structures both in the bulk and on the surface. In such a $\mathcal{C}$-WM, the  energies of the two Weyl points  are oppositely shifted  by an amount of $b_0$, 
 rendering  an electron-like FS (denoted by FS$_{+}$) and a hole-like FS (denoted by FS$_{-}$), as shown in Fig.~\ref{fig1}(b).
Owing to the $\mathcal{C}$-symmetry, the two FS's,  which are separated in momentum by $2\bm Q$, are well nested.  At low temperatures, particle-hole instabilities between FS$_\pm$ are induced by a repulsive interaction.
In particular, we consider
 three spin-density-wave orders (SDW$_{i}$, with spin indices $i=x,y,z$) and a charge-density-wave order (CDW), all with the same wavevector $2\bm Q$.
For an attractive interaction, uniform SC does {\it not} develop due to the anti-nesting of relevant FS regions. In this situation only two intra-FS pair-density-wave orders (PDW$_\pm$ for FS$_\pm$) with wavevectors $\pm 2\bm Q$ can develop. Thus, unlike many known cases, the leading weak-coupling instability in the particle-particle channel is unambiguously toward a PDW order.
We find that both PDW$_{\pm}$ and SDW$_z$ (the longitudinal SDW, where $z$ is parallel to $\bm{Q}$) gap out the full FS's, while the other orders leave gapless nodal lines on the FS's, as summarized  in Table \ref{table:node}. The nodal structures of the SDW, CDW and PDW orders are in sharp contrast with that for the uniform SC order in a doped WM~\cite{Lihaldane:2015aa}, which was found to have at least two point nodes. We found this distinction actually has a topological origin related to the Berry flux through the Weyl FS's. 
Since the density-wave orders that fully gap the FS's maximize the condensation energy, the leading instabilities are towards the SDW$_z$ for a repulsive interaction, and towards PDW$_\pm$ for an attractive interaction. We verify this via an explicit evaluation of the critical temperatures of all orders.

    \begin{figure}
 \includegraphics[width=1\columnwidth]{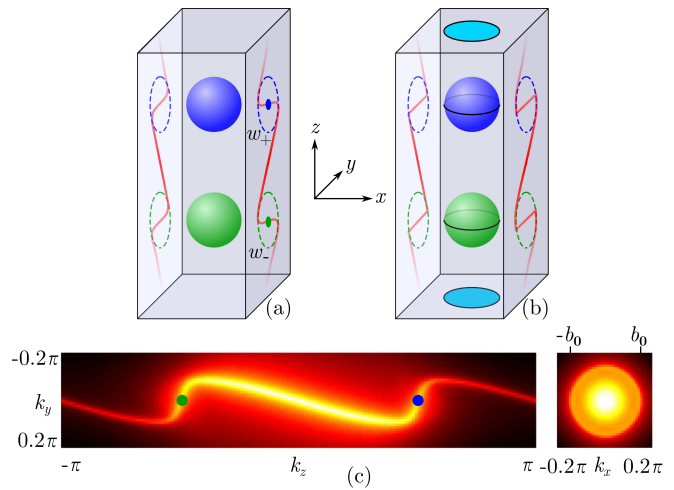}
 \caption{(Color online) Surface states in a density-wave ordered $\mathcal{C}$-WM. The positions of the original FS's are still shown for comparison. (a): The surface Fermi lines (red) of a SDW$_z$ state. On each surface they pass the projections (denoted by $w_{\pm}$) of the Weyl point  and extend beyond. 
  The case for the PDW state is similar. (b): The surface states of the CDW state. The bulk line nodes marked in solid lines project to in-gap drumhead bands on the $xy$-surfaces. (c): Simulation of the ARPES data on $yz$-surface for SDW$_z$ (left) and $xy$-surface for CDW (right). For simplicity we have set $2Q=\pi$, and the color intensity is at logarithmical scale. For the SDW$_z$ state, the spectral intensity of quasiparticles on the Fermi line becomes weaker outside the first folded BZ.}
     \label{fig2}
 \end{figure}
   
 We further analyze the fate of the surface states 
  in the presence of the density-wave orders.
 For the SDW$_z$ state, the bulk is fully gapped and becomes a weak topological insulator. This means that the Fermi arc on each $yz$-surface of the SDW$_z$ state has nowhere to terminate. We show that the surface Fermi line, which in the $\mathcal{C}$-WM state terminates on the $yz$-surface projections of FS's [Fig.~\ref{fig1}(b) and (c)], now traverses the full  surface $\bm k$-space [Fig.~\ref{fig2}(a)] in the  SDW$_z$ state. For angular-resolved photoemission spectroscopy (ARPES) measurements, the spectral intensity of the surface Fermi line is weaker beyond the first folded  Brillouin zones (BZ), due to vanishing quasiparticle spectral weights. Thus, the surface state  as seen by ARPES   still resembles an arc but no longer has sharp terminations, as shown in Fig.~\ref{fig2}(a) and (c).
  Moreover,  due to the $\mathcal{C}$-symmetry, the new surface Fermi line necessarily passes the projections of the Weyl points $w_{\pm}$ in Fig.~\ref{fig2}(a). 
 For the PDW state, the surface Fermi line is similar to that in the  SDW$_z$ case. Interestingly, {the zero modes at $w_{\pm}$ are  Majorana modes} as a result of  $\mathcal{C}$-symmetry in the Nambu space.    The surface Fermi line characterizes a 2D chiral Fermi liquid on each surface.  {These chiral surface states, together with the fully-gapped bulk, realizes a 3D anomalous Hall effect and support {dissipation-less} transport with a Hall conductivity $\sigma_H=e^2Q/h$ (Ref.\ \onlinecite{burkov_theta}) that is {\it independent} of the size of the material.} Near the onset of the  SDW$_z$ or PDW, the order parameter fluctuations around their expectation values are soft. As such, the coupling between the surface chiral fermions with the near-critical bosonic modes are expected to give rise to a surface chiral non-Fermi liquid. Several interesting properties of the chiral non-Fermi liquid have been explored in details by Sur and Lee \cite{Sur_lee2014}. We argue that our results in $\mathcal{C}$-WM provide a natural mechanism for realizing such a system.
 
Although the  CDW instability is not the leading one, it may emerge upon changing microscopic parameters, or simply externally introduced. Even when the CDW order is extrinsically induced, the topological band structure of the $\mathcal{C}$-WM still enforces the nodal line,  thereby resulting in a line-nodal semimetal~\cite{BHB2011}.
 We show that, besides the Fermi lines on $yz$-surfaces, there is one ``drumhead" band \cite{Chan:2015aa,bian_2016,Matsuura2013,Ramamurthy:2015aa} on each $xy$-surface with energies inside the CDW gap. The momentum range of the drumhead band in the 2D surface BZ  corresponds to the $xy$-surface projection of the nodal lines [Fig.~\ref{fig2}(b) and (c)]. Moreover, if the system has a $\mathcal{TI}$ symmetry (a product of an anti-unitary operation and spatial inversion)~\cite{Ramamurthy:2015aa}, the two drumhead bands on the opposite surfaces are degenerate. 
 As a result, there exists a symmetry-protected topological charge density $j^0=\pm eb_0^2/(8\pi)$ on each $xy$-surface.

The remainder of this paper is organized as follows. In Sec.~\ref{section:model}, we study the bulk and surface states of a $\mathcal{C}$-Weyl metal. 
In Sec.~\ref{section:bulk_of_cdw_sdw}, we study the CDW and SDW instabilities with a repulsive interaction and address their respective nodal structures in the bulk. In Sec.~\ref{section:bulk_of_pdw}, We analyze the PDW instability in the presence of a weak attractive interaction, and for comparison, we show the uniform SC does not emerge.  In Sec.~\ref{section:LH}, we perform a systematic analysis of the structure of CDW, SDW, and PDW using a Berry flux argument, in comparison with an eariler study~\cite{Lihaldane:2015aa} done for the uniform SC order.  
 In Sec.~\ref{section:chiral_sdw_pdw} we present the structure of  the chiral surface states of the SDW and PDW ordered bulk. Sec.~\ref{section:cdw} is devoted to the drumhead surface states of the CDW bulk and its topological properties.
 Conclusions and possible relations to experiments are discussed in Sec.~\ref{section:conclusion}.

\section{Model of a $\mathcal{C}$-Weyl metal and Instabilities}\label{section:model}

\subsection{Lattice model of a $\mathcal{C}$-Weyl metal}
 We begin with a simple two-band lattice Hamiltonian $H=\sum_{\bm k} \psi^\dagger({\bm k}) h({\bm k}) \psi({\bm k})$ for the $\mathcal{C}$-WM:
\begin{align}
\!\!h({\bm k})=&\sin k_x \sigma^x+\sin k_y \sigma^y+ \left(\cos k_z-\cos  {Q}\right)\sigma^z/\sin Q\nonumber\\
&- (2-\cos k_x-\cos k_y)\sigma^z+b_0 {\sin k_z}/{ \sin{Q}}\,,
\label{lat}
 \end{align}
 where $\sigma^{x,y,z}$ are Pauli matrices for the spin. 
 The first four terms give rise to the two Weyl nodes $W_\pm$ at the BZ points $\pm \bm Q=   (0,0, \pm Q)$ with chiralities $\chi=\mp 1$. The Chern number $C=1$ for $\bm k$-space slices with $|k_z|<Q$, and $C=0$ for those with $|k_z|>Q$.
  The model breaks time-reversal symmetry, and  inversion symmetry, but preserves instead a $\mathcal{C}$-symmetry:
 $\mathcal{C}h^T (-{\bm k}) \mathcal{C}^{-1}=-h(\bm k),$
 where ${\mathcal C}=\sigma^x$.  The last term in Eq.~(\ref{lat}) shifts the energies of the two Weyl points by $\pm b_0$, and is known to give rise to the chiral magnetic effect~\cite{aji2012,12a,Son_2012,burkov_theta,Grushin_2012,hosur2013,goswami2013}. A similar two-band model has been shown by Burkov and Balents~\cite{burkov_balents2011} to emerge from a topological insulator -- trivial insulator (TI-SI) heterostructure, where
 the last term of Eq.~(\ref{lat}) is generated by including the spin-orbit coupling between the TI-SI interfaces~\cite{burkov2012}. Such a term that is odd in momentum without a Pauli matrix structure was also analyzed in the context of the so-called ``type-II" Weyl semimetals~\cite{Soluyanov:2015aa,type2a,type2b,type2c,type2d,type2e,Liang:2016aa,xuzhangzhang2015}, where both electron and hole FS's exist. It has also been recently showed that the topology of the hole and electron FS's can also be induced by pressure~\cite{bhyanpressure}.
  
 For the parameter region $b_0\ll Q$ that we focus on below, the low energy fermions can be described by expanding Eq.\ \eqref{lat} in $\bm k$-space around $W_\pm$. The resulting  energies are given by $E_\lambda\approx\lambda\sqrt{p_x^2+p_y^2+p_z^2}+b_0\tau^z$, where $\lambda=\pm1$,    $\bm p$ is the momentum deviation from $w_{\pm}$, and
   $\tau^z=\pm 1$ distinguishes $w_{\pm}$.  
There are two Weyl Fermi surfaces, each corresponding to $\tau^z=-\lambda=\pm 1$. For $\lambda=-\tau^z=1(-1)$ the Fermi velocity points outwards (inwards) from the encapsulated Weyl point   and the corresponding FS is electron-like (hole-like).   We denote the two FS's as FS$_\pm$. Both FS$_\pm$ are spin textured, as schematically shown in Fig.\ \ref{fig1}(b).  
The two FS's are nested, as they have identical shape and opposite Fermi velocity orientations. Note that this nesting actually does not require the exact form of the lattice model (\ref{lat}), but rather is a general result of the $\mathcal{C}$-symmetry and linear expansion of the fermionic dispersion around Weyl points. Generically, the dispersion around $W_+$ is given by $h_+=b_0+k_{i}A_{ij}\sigma^j$, ($i,j=x,y,z$). According to the $\mathcal{C}$-symmetry, the dispersion around $W_-$ is then $h_-=-b_0 - k_{i} A_{ij} \tilde\sigma^j$, [$\tilde\sigma^i\equiv \mathcal{C}(\sigma^i)^T\mathcal{C}^{-1}$]. It is easy to verify the resulting FS$_\pm$ are nested.

\subsection{Fermi arcs on the surface of a $\mathcal{C}$-Weyl metal}

The surface Fermi arcs on the two opposite $yz$-surfaces,  terminate on the projections of FS$_\pm$ [Fig.~\ref{fig1}(b),(c)]. The position of each Fermi arc is  given by $k_y\approx\pm b_0 k_z/\sin Q$  in the small $b_0$ limit.

More concretely, we begin with Eq.~(\ref{lat}), 
For a given $|k_z|<Q$, it is easy to see that the lower band of the 2D momentum space slice $h_{2D}(k_x,k_y)\equiv h(k_x,k_y,k_z)$ has Chern number $C=1$, due to a skyrmion configuration of the spin over the 2D Brillouin zone (BZ). As a result of the Chern number, if we consider open boundaries in $x$ direction,  there exist chiral modes on the two edges.
The dispersion of the edge modes of each $k_z$ slice can be obtained in the continuum limit by standard procedures. In the continuum limit,
\begin{align}
h_{\rm cont}(\bm k)=k_x\sigma^x+k_y\sigma^y+m_z\sigma^z+b_z,
\end{align}
where $m_z=({\cos k_z -\cos Q})/{\sin Q}>0$ and $b_z=b_0\sin k_z/\sin Q$.
A boundary in $x$ direction  {between the $\mathcal{C}$-WM and the vaccuum (which is equivalent to a trivial insulator)} can be modeled by a domain wall of $m_z$, namely $m_z(x)=-m_z\sgn(x)$, where $x<0$ is the $\mathcal{C}$-WM side. The continuum Hamiltonian becomes $h_{\rm cont}(x,k_y)=-i\partial_x \sigma^x+m_z(x)\sigma^z+k_y\sigma^y+b_z$. The eigenfunction of this Hamiltonian  for the surface state is $\psi(x)=\psi_0\exp(\int^x m_z(x')dx')$ with $\sigma^y\psi^0=-\psi^0$. The energy eigenfunction, on the other hand, is given by
\begin{align}
E_{\rm surf}=-k_y+b_z=-k_y+b_0\sin k_z/\sin Q.
\end{align}
It is straightforward to obtain that for the opposite surface where $m_z(x)=m_z\sgn(x)$,  $E'_{\rm surf}=k_y+b_z=k_y+b_0\sin k_z/\sin Q$.
The position of the surface Fermi arc is by definition given by $E_{\rm surf}=0$, and therefore for the fermi arcs localized on the two surfaces are given by, in the continuum limit $k_y\ll Q$,
\begin{align}
k_y=\pm b_0 k_z/\sin Q.
\end{align}
This result is illustrated in Fig.~\ref{fig1}.  In Fig.~\ref{fig1}(c) as well as Fig.~\ref{fig2}(c), we have set $2Q=\pi$, $k_y=0$, $b_0= 0.3$ (in units where $v_F=1$) and $\rho=0.2$ and simulated the ARPES data.

\section{ Charge- and spin-density-wave orders in particle-hole channels} \label{section:bulk_of_cdw_sdw}
\subsection{Projective form factors and nodal structures}
 Since FS$_\pm$ are well nested, with  some small repulsive interaction~\cite{burkov_theta,nandkishore2014}, e.g., a screened Coulomb interaction, the $\mathcal{C}$-WM is intrinsically unstable in the particle-hole channel. Specifically, we consider density-wave orders with wavevector $2\bm Q$ that couple to fermions via terms 
 \begin{align}
\mathcal{H}_\rho&=\rho~\psi^\dagger(\bm p+\bm Q) \psi(\bm p-\bm Q)+h.c., \label{dw0}\\
 \mathcal{H}_M^i&= M^i\psi^\dagger\!(\bm p+\bm Q) \sigma^i \psi(\bm p-\bm Q)+h.c.,
 \label{dw}
 \end{align} 
 where   $\rho$ represents  the amplitude of CDW order, and $M^i$ represent the  amplitude of  spin-density-wave orders (SDW$_i$) with spin orientation $i=x,y,z$. These density-wave orders has been introduced in WMs \cite{burkov_theta,franz2013} and WSMs~\cite{Wang2013axion,roy_sau2015,wclee2016,you_cho_hughes2016} in the study of  chiral anomaly \cite{aji2012,12a,Son_2012,Grushin_2012,burkov_theta,hosur2013,goswami2013}. However, a detailed analysis of their intrinsic structures are still lacking.
  Below we provide a systematic analysis of the instabilities towards these orders  in the $\mathcal{C}$-WM. 
  
Even though $M^i$ and $\rho$ are constants (as we will show via detailed calculations in next Subsection) in momentum, they project onto the spin-polarized FS$_{\pm}$ with different form factors. We dub the form factors on spin-textured FS's as ``projective form factors".
More specifically, we express the spin polarizations of the low energy fermions $c_{\pm}^\dagger(\bm p)$ on FS$_\pm$ as 
$c_{\pm}^\dagger(\bm p)=\sum_\alpha \xi_{\pm,\alpha}(\hat{\bm p}) \psi_\alpha^\dagger(\bm p \pm  \bm Q)$, where $\alpha$'s are spin indices. The spinor part of the Bloch  wave function is given by  
\begin{align}
\xi(\bm p \pm \bm Q)\equiv \xi_{\pm} (\hat {\bm p})=\[\sin \frac{\theta}{2},~ \pm \cos\frac{\theta}{2} \exp(-i\varphi) \]^T,\label{sup:eq3}
\end{align}
where $(\theta,\varphi)$ are defined as spherical coordinates via $\bm p=(b_0\sin\theta\cos\varphi , b_0\sin\theta\sin\varphi,  b_0\cos\theta)$. The CDW and SDW Hamiltonians project onto low energy fermions $c_\pm$ as
$\bar{\mathcal{H}}_\rho= \rho\chi_\rho(\hat{\bm p}) c_+^\dagger(\bm p) c_-(\bm p)+h.c.\,$ and $
\bar{\mathcal{H}}_M^i = M^i\chi_M^i(\hat{\bm p})  c_+^\dagger(\bm p) c_-(\bm p)+h.c.
  $. After simple math, we find that  the {projective form factors} are given by:
  \begin{align}
  &\chi_{\rho}(\hat{\bm p})=\xi^\dagger_{+}(\hat{\bm p})\xi_{-}(\hat{\bm p})=-\cos\theta\,, \label{projective_rho}\\
 & \chi_{M}^x(\hat{\bm p})=\xi^\dagger_{+}(\hat{\bm p})\sigma^x\xi_{-}(\hat{\bm p})=i\sin\theta\sin\varphi\,, \label{projective_mx} \\
 &\chi_{M}^y(\hat{\bm p})=\xi^\dagger_{+}(\hat{\bm p})\sigma^y\xi_{-}(\hat{\bm p})=i\sin\theta\cos\varphi\,,\label{projective_my} \\
 &\chi_{M}^z(\hat{\bm p})=\xi^\dagger_{+}(\hat{\bm p})\sigma^z\xi_{-}(\hat{\bm p})=1\,.\label{projective_mz}
 \end{align}
 It is clear that  SDW$_z$   gaps out the full FS, while CDW and SDW$_{x,y}$ have line nodes~\cite{footnote}. We illustrate their nodal structures in Table \ref{table:node}.

   \begin{table}
\centering
\caption{Bulk nodal structures of density-wave ordered Weyl metals. The positions of the nodal lines are marked by solid lines around an original spherical FS. The $x,y,z$ directions are the same as in Fig.~\ref{fig1}.}
\label{table:node} 
\begin{ruledtabular}
 \begin{tabular}{ccccc}
CDW & SDW$_x$ & SDW$_y$ & SDW$_z$ & PDW$_\pm$\\
  \hline
\begin{minipage}[c]{0.15\columnwidth}{\includegraphics[width=\columnwidth]{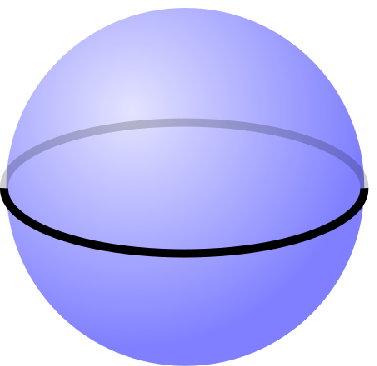}}\end{minipage}&
\begin{minipage}[c]{0.15\columnwidth}{\vspace{0.5mm}\includegraphics[width=\columnwidth]{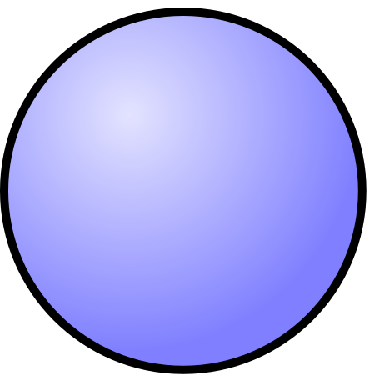}}\end{minipage}&
\begin{minipage}[c]{0.15\columnwidth}{\includegraphics[width=\columnwidth]{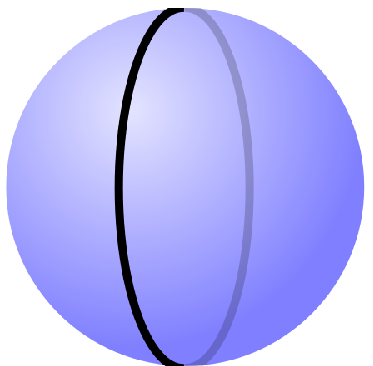}}\end{minipage}&
\begin{minipage}[c]{0.15\columnwidth}{Full gap}\end{minipage} &
\begin{minipage}[c]{0.15\columnwidth}{Full gap}\end{minipage}  \end{tabular}
 \end{ruledtabular}
 \end{table}

  \subsection{Linearalized gap equations and critical temperatures} 
 The nodal structures of the different orders in the particle-hole channel in Eqs.~(\ref{dw0},\ref{dw}) distinguish their respective critical temperatures. For a repulsive interaction,  SDW$_z$ (i.e., the longitudinal SDW) has the highest critical temperatures among SDW and CDW orders, heuristically because it gaps out the full FS's (see Table \ref{table:node}) and maximizes
  the condensation energy.  
  
    Below we perform a more rigorous analysis by directly solving the linearized gap equations for the SDW and CDW orders.
   We analyze the longtitudinal SDW$_z$ order first. For the constant interaction repulsive $V$, we take and later verify the ansatz $M^z(\bm p)=M^z$, and the linear gap equation is given by
\begin{align}
M^z\sigma^z_{\alpha\beta}
=&-VT\sum_{\omega_m}\int \frac{d^3\bm p}{(2\pi)^3}G_{\alpha\gamma}(\omega_m, \bm p+\bm Q)\nonumber\\
&\times G_{\delta\beta} (\omega_m, \bm p-\bm Q) M^z \sigma^z_{\gamma\delta},
\label{sup:gap}
\end{align}
where $G_{\alpha\beta}(\omega_m, \bm k)=\langle \mathcal{T} \psi_\alpha(\omega_m, \bm k) \psi_\beta^\dagger(\omega_m, \bm k) \rangle $ is the spin-dependent fermionic Green's function. $\omega_m$ is the fermionic Matsubara frequency. In practice for a given $\bm k$, only one of the two spin orientations of the fermion is at low energy and has a Fermi surface (FS)~\cite{wang_new}.  This low-energy spin orientation is given by the spinor in Eq.~(\ref{sup:eq3}). Since the instabilities are dominantly driven by the nesting of the FS$_\pm$, we can safely project out the high energy spin polarization. Thus,
\begin{align}
G_{\alpha\beta}(\omega_m, \bm p\pm \bm Q)\equiv G_{\alpha\beta}^{\pm}(\omega_m, \bm p)=&-\frac{\xi_{\pm,\alpha}(\hat {\bm p})\xi_{\pm,\beta}^*(\hat {\bm p})}{i\omega_m\mp\epsilon(\bm p)},
\label{sup:G}
\end{align} 
where $\xi_{\pm}(\bm p)\xi^*_{\pm}(\bm p)$ is the spin projection operator, and $\pm\epsilon(\bm p)$ is the fermionic dispersion for the electron (hole) FS. Plugging Eq.\ \eqref{sup:G} into Eq.\ \eqref{sup:gap}, we obtain
\begin{align}
M^z\sigma^z_{\alpha\beta}=&VN(0)T\sum_{\omega_m}\int \frac{d\epsilon}{\omega_m^2+\epsilon^2}\int \frac{\sin\theta d\theta d\varphi}{4\pi}\nonumber\\
&\times\xi_{+,\alpha}(\hat {\bm p})\xi_{-,\beta}^*(\hat {\bm p})\chi_M^z(\hat {\bm p}) M^z,
\label{sup:gap2}
\end{align}
where $N(0)$ is the density of states on FS$_\pm$, and, $\chi_M^z(\hat {\bm p})=1$ is precisely the projective form factor given in Eq.~(\ref{projective_mz}). The angular integral in Eq.\ \eqref{sup:gap2} can be carried out as
\begin{align}
M^z\sigma^z=&VN(0)T\sum_{\omega_m}\int \frac{d\epsilon}{\omega_m^2+\epsilon^2}\int \frac{\sin\theta d\theta d\varphi}{8\pi}\nonumber\\
&\times\[\begin{array}{cc} -\cos{\theta}+1&-\sin{\theta}e^{i\varphi}\\ \sin{\theta}e^{-i\varphi}& -\cos{\theta}-1 \end{array}\]
M^z,\nonumber\\
=&\frac{VN(0)}{2}T\sum_{\omega_m}\int \frac{d\epsilon}{\omega_m^2+\epsilon^2}M^z\sigma^z.
\label{sup:gap3}
\end{align}
From Eq.\ \eqref{sup:gap3}, we see that the Pauli matrix structures from the l.h.s.\ and the r.h.s.\ indeed match. The factor 2 in the denominator is the manifestation of the fact that only one spin orientation of the fermions contributes to the instability.
 The integral and summation over $\epsilon$ and $\omega_m$ produce the usual logarithmical divergence as $T\to 0$. By standard procedures, we obtain the critical temperature for $M^z$ to have a nontrivial solution is
\begin{align}
T_{c1}=\Lambda e^{-\frac{2}{N(0)V}}\equiv \Lambda e^{-{1}/{g}},
\label{sup:tc1}
\end{align}
where we have defined the dimensionless repulsive coupling $g\equiv N(0)V/2$.

The critical temperatures for other particle-hole instabilities can be similarly obtained. For the SDW$_x$ order  $M^x(\bm p)=M^x$,
\begin{align}
M^x\sigma^x_{\alpha\beta}=&VN(0)T\sum_{\omega_m}\int \frac{d\epsilon}{\omega_m^2+\epsilon^2}\int \frac{\sin\theta d\theta d\varphi}{4\pi}\nonumber\\
&\times\xi_{+,\alpha}(\hat {\bm p})\xi_{-,\beta}^*(\hat {\bm p})\chi_M^x(\hat {\bm p}) M^x\,
\label{sup:gap4}
\end{align}
and the only difference with Eq.\ \eqref{sup:gap2} lies in the projective form factor $\chi_M^x(\hat {\bm p})$ is given in Eq.~(\ref{projective_mx}). An angular integration similar to Eq.\ \eqref{sup:gap3} yields
\begin{align}
M^x\sigma^x=&VN(0)T\sum_{\omega_m}\int \frac{d\epsilon}{\omega_m^2+\epsilon^2}\int \frac{\sin\theta d\theta d\varphi}{8\pi}\nonumber\\
&\times\[\begin{array}{cc} -\cos{\theta}+1&-\sin{\theta}e^{i\varphi}\\ \sin{\theta}e^{-i\varphi}& -\cos{\theta}-1 \end{array}\](i\sin\theta \sin\varphi)
M^x,\nonumber\\
=&\frac{VN(0)}{6}T\sum_{\omega_m}\int \frac{d\epsilon}{\omega_m^2+\epsilon^2}M^x\sigma^x.
\label{sup:gap5}
\end{align}
We indeed find that the Pauli matrix structures match for both sides and the critical temperature is given by
\begin{align}
T_{c2}=\Lambda e^{-3/g}.
\label{sup:tc2}
\end{align}
It is straightforward to repeat the same procedure for CDW order $\rho$ and SDW$_y$ order $M^y$, and the only difference is that their projective form factors  
$\chi_\rho=-\cos\theta$ and $\chi_{M}^y$ are given in Eq.~(\ref{projective_rho}) and (\ref{projective_my}) respectively. After performing the angular integral over the FS, we find that their critical temperatures are also given by 
 \begin{align}
 T_{c2}=\Lambda \exp({-3/g}).
 \end{align}
       

The result of our comparative analysis based on nodal structures of density-wave orders in the $\mathcal{C}$-WM can be formally extrapolated to the WSM limit ($b_0\to 0$), \emph{even if} both FS$_\pm$ and the line nodes reduce to points. Indeed, a renormalization group (RG) analysis in a WSM 
 \cite{nandkishore2014} has also identified the longitudinal SDW as the leading instability.  However, due to the vanishing  density of states at the Weyl points in the WSM limit,
  the interaction in a WSM has to be strong enough in order to induce density-wave instabilities, and 
  strictly speaking, the weak coupling RG method they adopted becomes uncontrolled in this regime. Besides, the subleading SDW and CDW instabilities identified in our analysis are not present in their analysis.
    
  \section{Pair-density wave orders in particle-particle channels}\label{section:bulk_of_pdw}
With some attractive interaction, e.g., from electron-phonon coupling, the $\mathcal{C}$-WM also has instabilities in the particle-particle channel. Unlike a case with either two electron FS's or two hole FS's \cite{Bednik2015,LuTanaka2015,Lihaldane:2015aa,Wang:2016aa,Cho_weyl_2012}, there is no instability towards a spatially uniform SC in the $\mathcal{C}$-WM, because the fermions that are paired for uniform SC have {\it opposite} energies. Nonetheless, there do exist intra-FS pairing instabilities~\cite{Bednik2015,Cho_weyl_2012}.
  Since each FS is centered around a nonzero momentum $\pm\bm Q$, such pairing results in a non-uniform superconducting state with finite pairing momentum $\pm 2{\bm Q}$, i.e., the PDW$_\pm$ state~\cite{Bednik2015,LuTanaka2015,Lihaldane:2015aa,Wang:2016aa,Cho_weyl_2012}. The corresponding PDW$_{\pm}$ order parameters in FS$_\pm$ are given by $\phi_{\pm}\sim\langle\psi^\dagger(\bm p\pm\bm Q)(i\sigma^y)\psi^\dagger(-\bm p\pm\bm Q)\rangle$. 
     
Below we solve the linear gap equation for PDW order and obtain its critical temperature.  The analysis of the PDW order for the $\mathcal{C}$-WM is analogous to that for a doped WM~\cite{Cho_weyl_2012,Bednik2015}, and for comparison with SDW and CDW orders we present the detailed calculation here. Since we are mainly concerned with the projective form factor of the order parameter, we consider the simple density-density interaction $-U<0$ that is attractive and independent of momentum transfer.
We focus on the PDW$_+$ order, and it is trivial to check that the onset temperature for PDW$_-$ is identical to that for PDW$_+$. 
The gap equation for PDW$_+$ order parameter $\phi_+$ is
\begin{align}
&\phi_+(i\sigma^y)_{\alpha\beta}\nonumber\\
=&UT\sum_{\omega_m}\int \frac{d^3\bm p}{(2\pi)^3}G^+_{\alpha\gamma}(\omega_m, \bm p)G^+_{\beta\delta} (-\omega_m, -\bm p) \phi_+ (i\sigma^y)_{\gamma\delta}\nonumber\\
=&UN(0)T\sum_{\omega_m}\int \frac{d\epsilon}{\omega_m^2+\epsilon^2}\int \frac{\sin\theta d\theta d\varphi}{4\pi}\nonumber\\
&\times\xi_{+,\alpha}(\hat {\bm p})\xi_{+,\beta}(-\hat {\bm p})\chi_{\phi_{+}}(\hat {\bm p}) \phi_+.
\label{sup:gap6}
\end{align}
Note that in the last line we have automatically obtained the projective form factor of PDW$_+$ as  $\chi_{\phi_+}(\hat{\bm p})\equiv\xi_{+}^{\dagger}(\hat {\bm p})(i\sigma^y) \xi_{+}^*(-\hat {\bm p})=-\exp(i\varphi)$. Other than the north and south poles where $\varphi$ is ill-defined, the PDW$_{+}$ order gaps out the full FS. (After a gauge transformation the singular points at the two poles can be removed.) Hence, the PDW order parameter gaps out a full FS \cite{Cho_weyl_2012,Bednik2015}, which is included in Table \ref{table:node}.

Further plugging in the expression for $\xi_{+}(\pm \hat{\bm p})$, the angular integral is performed as
\begin{align}
\phi_+ \sigma^y=&UN(0)T\sum_{\omega_m}\int \frac{d\epsilon}{\omega_m^2+\epsilon^2}\int \frac{\sin\theta d\theta d\varphi}{8\pi}\nonumber\\
&\times\[\begin{array}{cc} -\sin\theta e^{i\varphi}&-\cos\theta+1 \\ -\cos\theta-1 & \sin\theta e^{-i\varphi}  \end{array}\] \phi_+\nonumber\\
=&\frac{UN(0)}{2}T\sum_{\omega_m}\int \frac{d\epsilon}{\omega_m^2+\epsilon^2} \phi_+ \sigma^y\,.
\end{align}
Again, the factor 2 in the denominator is due to the fact that only one of the two spin directions of a fermion participates the pairing.  
The critical temperature of the PDW order is given by
\begin{align}
T_{c3}=\Lambda e^{-\frac{2}{N(0)U}}\equiv \Lambda e^{-{1}/{g'}}\,.
\end{align}
 
{It is also instructive to analyze the uniform SC order which does not develop. The linear gap equation for the SC order parameter $\Delta$ is
\begin{align}
&\Delta(i\sigma^y)_{\alpha\beta}\nonumber\\
=&UT\sum_{\omega_m}\int \frac{d^3\bm p}{(2\pi)^3}G^+_{\alpha\gamma}(\omega_m, \bm p)G^-_{\beta\delta} (-\omega_m, -\bm p) \Delta (i\sigma^y)_{\gamma\delta}\nonumber\\
=&-UN(0)T\sum_{\omega_m}\int \frac{d\epsilon}{(i\omega_m-\epsilon)^2}\int \frac{\sin\theta d\theta d\varphi}{4\pi}\nonumber\\
&\times\xi_{+,\alpha}(\hat {\bm p})\xi_{+,\beta}(-\hat {\bm p})\chi_{\Delta}(\hat {\bm p}) \Delta.
\label{sup:gap6}
\end{align}
Even before the angular integral, one immediately finds that the integral over $\epsilon$ vanishes, due to the double pole $\epsilon=i\omega_m$ in the complex plane. Hence for a weak constant interaction $U$, the uniform SC does not develop.

\section{Topological properties of the projective form factors}\label{section:LH}

The fact that the CDW, SDW$_{i}$ and PDW orders have smooth projective form factors $\chi$'s (i.e., no vortices) over the full FS's is in sharp contrast with the case of uniform SC in a doped WM~\cite{Bednik2015,LuTanaka2015,Lihaldane:2015aa,Wang:2016aa,Cho_weyl_2012}. As pointed out by Li and Haldane \cite{Lihaldane:2015aa}, the gap for uniform SC of a doped WM has to have at least two {topologically robust} point nodes due to a $4\pi$ Berry flux carried by the pairing field through a FS. Below we show that the Berry fluxes carried by the CDW and SDW fields through each FS vanish, even though the fermionic field $\psi$ itself carries a nonzero Berry flux. Therefore, the resulting form factors in our case are allowed to be non-singular (and even a constant for SDW$_z$). 
In this sense, the nodal lines for the CDW and SDW$_{x,y}$ are not topologically robust alone. However, they can be protected by an additional $\mathcal {TI}$ symmetry~\cite{Ramamurthy:2015aa}. We find that such a symmetry indeed  can be present for the CDW state (see Sec. \ref{section:cdw}). 


In the $\mathcal{C}$-WM, the Berry  flux of the fermion wave function $\xi_{\pm}$ through FS$_\pm$ is given by 
\begin{align}
\Phi_\pm\equiv\oint_{\rm{FS}_{\pm}} d\bm p\cdot \nabla_{\bm p}\times \mathcal{A}_{\pm}(\bm p)= 2\pi,
\end{align}
 where the Berry connection $\mathcal{A}_{\pm}(\bm p)\equiv\xi_{\pm}^\dagger(\bm p) (i\nabla_{\bm p}) \xi_{\pm}(\bm p)$. This result is well-known, and accounts for the fact that $\xi_\pm$ cannot be well-defined everywhere on the FS$_\pm$.
The fact that $\Phi_+=\Phi_-$ is a combination of the following two effects -- (i) that the two Weyl points of the original WSM state are of opposite chiralities, and (ii) that the FS$_\pm$ are formed by upper and lower branches of the energy eigenvalues respectively. 

The Berry   flux can be generalized to fermionic bilinear operators~\cite{nagaosa}. Particularly in the particle-hole channel between FS$_\pm$, the Berry connection for the CDW and SDW operators are given by:
\begin{align}
\mathcal{A}_{ph}(\bm p)=\mathcal{A}_{\pm}(\bm p)-\mathcal{A}_{\mp}(\bm p),
\end{align}
 where the relative minus sign comes from fact that creation and  annihilation operators rotate oppositely under a gauge tranformation. In the particle-particle channel for PDW$_\pm$, the Berry connection is 
 \begin{align}
 \mathcal{A}_{pp}(\bm p)=\mathcal{A}_{\pm}(\bm p)+\mathcal{A}_{\pm}(-\bm p).
 \end{align} 
 The Berry flux carried by the CDW and SDW order parameters through a FS, say FS$_{+}$, satisfies 
 \begin{align}
 \Phi_{ph}\equiv&\oint_{\rm{FS}_{+}} d\bm p\cdot \nabla_{\bm p}\times \mathcal{A}_{ph}(\bm p)\nonumber\\
 =&\oint_{\rm{FS}_{+}} d\bm p\cdot \nabla_{\bm p}\times[\mathcal{A}_{+}(\bm p)-\mathcal{A}_{-}(\bm p)]\nonumber\\
 =&\Phi_{+}-\Phi_{-}=0,
 \label{sup:eq18}
 \end{align}
 while the Berry flux for PDW orders, for example $\phi_+$ satisfies
  \begin{align}
\Phi_{pp}\equiv&\oint_{\rm{FS}_{+}} d\bm p\cdot \nabla_{\bm p}\times \mathcal{A}_{pp}(\bm p)\nonumber\\
=&\oint_{\rm{FS}_{+}} d\bm p\cdot \nabla_{\bm p}\times[\mathcal{A}_{+}(\bm p)+\mathcal{A}_{+}(-\bm p)]\nonumber\\
=&\Phi_{+}-\Phi_{+}=0.
\label{sup:eq19}
  \end{align}
The vanishing of the Berry fluxes indicates that the projective form factor $\chi$'s (analog of wave function for the fermions) for CDW, SDW, and PDW order parameters can be smooth and nonsingular everywhere on the FS. Particularly, for SDW$_z$ and PDW$_{\pm}$, the projective form factors are a constant.
This is in contrast with the SC case in a doped WM, where the Berry flux of the  SC order parameter was found~\cite{Lihaldane:2015aa} to be $4\pi$, and the resulting SC gap necessarily has point nodes\cite{Cho_weyl_2012,LuTanaka2015,Lihaldane:2015aa}.

{From this argument, the line nodes for CDW and SDW$_{x,y}$ orders are not topologically robust, and can be eliminated by perturbations. For example, it is straightforward to check that the CDW line node induced by an imaginary order parameter $\rho$ can be eliminated by introducing a small SDW$_z$ order parameter $M^z$ that is real. However, as we show in Sec.\ \ref{section:cdw}, the line nodes can be protected by a $\mathcal{TI}$ symmetry~\cite{Ramamurthy:2015aa}.}

 \section{ Chiral surface states of  the SDW/PDW   bulk}\label{section:chiral_sdw_pdw}

\subsection{Main results}
 In the SDW$_z$ state, the bulk becomes fully gapped. Therefore, the Chern numbers of all $k_z$ slices have to be the same, resulting in a 3D weak topological insulator formed by  stacking Chern insulators with $C=1$ for all $k_z$'s. Meanwhile, the surface Fermi arc, which in the $\mathcal{C}$-WM state used to terminate at the FS [Fig.\ \ref{fig1}(b)], cannot terminate anywhere, and necessarily traverses the full surface BZ. On a $yz$-surface, a Fermi arc portion that is far away from both FS$_\pm$ in the metallic state is not affected dramatically by $M^z$, since the density-wave order parameter concentrates around FS$_\pm$. For momenta close to FS$_\pm$, the position of the Fermi line is strongly modified.  Particularly,  {we find that, as a result of a $\mathcal{C}$-symmetry inherited from the metallic state,} the Fermi line in the SDW$_z$ state passes the $w_\pm$ points, i.e., $k_y=0, k_z=\pm Q$. 
This argument is further verified  by numerical calculation of  $yz$-surface states and the surface ARPES intensity profile [see Fig.~\ref{fig2}(c)]. 
The surface Fermi line in the PDW state is similar to the SDW case. Particularly, the Fermi line also passes  $w_{\pm}$ due to a $\mathcal{C}$-symmetry. However, this $\mathcal{C}$-symmetry is not inherited from the original metallic model (\ref{lat}) but from the superconducting Nambu representation. As a result, for the PDW state, the zero modes at $w_{\pm}$ are actually Majorana modes.

 In summary, we fixed the behavior of the surface Fermi line in the SDW$_z$ and PDW$_\pm$ states with two main arguments -- (a) since the bulk is fully gapped, the Chern number $C$'s for all $k_z$ slices are necessarily the same, and (b) for slices at $k_z=\pm Q$, there exists a particle-hole $\mathcal C$-symmetry for the 2D Hamiltonian. As a consequence, for both cases, the reconstructed surface Fermi line traverses the full $\bm k$-space, and passes the $k_y=0$ for $k_z=\pm Q$, i.e. the projections $w_{\pm}$ of the bulk Weyl points $W_{\pm}$. 
 
 \subsection{Details of analysis}
 In the following, we provide a detailed analysis on these arguments. 
For the SDW$_z$ case, we  consider the simplest ``period-two" case with $2Q=\pi$, and it is trivial to generalize to other values of $Q$. The single-particle Hamiltonian at $k_z=\pm Q=\pm \pi/2$  with $M^z$ term is,
\begin{align}
h_M^z(k_x, \!k_y)\!=\!&\sin k_x \sigma^x\!+\!\sin k_y \sigma^y\!-\!(2\!-\!\cos k_x\!-\!\cos k_y)\sigma^z\nonumber\\
&+b_0s^z+M^z\sigma^z s^x\,,
\label{k=q}
\end{align}
where the Pauli matrix $s^{z}$ distinguishes $k_z=\pm Q$. Similar to the original $\mathcal C$-WM, Eq.\ \eqref{k=q} has a $\mathcal{C}$-symmetry with a symmetry operation given by $\mathcal{C}'=\sigma^x s^x$, for which
\begin{align}
{\mathcal C}'(h_{M}^z)^T(-k_x,-k_y)\mathcal{C}'=h_{M}^z(k_x,k_y).
\end{align}
It is straightforward to see that this $\mathcal{C}'$ operator derives from the $\mathcal{C}$ operator for the original $\mathcal{C}$-WM Hamiltonian [Eq.~(\ref{lat}) with $k_z=\pm \pi/2$]. The $s^x$ operator takes $k_z\to-k_z$. However, such an ``inherited" $\mathcal{C}$-symmetry is not present for other $k_z$ slices, as under $s^x$, $k_z$ generally becomes $k_z \mp 2Q\neq -k_z$ for $k_z\neq \pm Q$.

This 2D Hamiltonian has $C=1$ and supports chiral edge modes. Due to the $\mathcal{C}$-symmetry the chiral modes must pass zero energy at the particle-hole symmetric momenta $k_y=0 {~\rm or~} \pi$~\cite{Qi2008}. Given that the original Fermi arc is at small $k_y$'s, we expect that the surface Fermi line  passes $k_y=0, k_z=\pm Q$, i.e., $w_{\pm}$ points.  This indeed is confirmed by numerics (see Fig. \ref{fig2}).

We move to the PDW case and invoke the similar arguments. We show below that for $k_z=\pm Q$ there also exists a $\mathcal{C}$-symmetry, although this $\mathcal{C}$-symmetry has a distinct nature from that for the SDW$_z$ case. To see this, we set $k_z=Q$ and go to the Nambu space,
\begin{align}
\!h_{\phi_+}(k_x,k_y)=&\sin k_x \sigma^x + \sin k_y \sigma^y \tau^z\nonumber\\
& -(2 - \cos k_x - \cos k_y)\sigma^z \tau^z \nonumber\\
&+ b_0 \tau^z - \phi_+ \sigma^y \tau^y,
\label{sup:hphi}
\end{align}
where we have neglected the $\phi_-$, since it concentrates on the other FS centered at $k_z=-Q$.
Under the operation  $\mathcal{C''}=\tau^x$, this Hamiltonian satisfies
\begin{align}
{\mathcal C}''h_{\phi_+}^T(-k_x,-k_y)\mathcal{C}''=h_{\phi_+}(k_x,k_y),
\end{align}
thus $\mathcal{C}''$ defines a particle-hole symmetry operation for Eq.\ \eqref{sup:hphi}. By the same reasoning, the reconstructed surface Fermi line in the PDW case passes the $w_{\pm}$ points at 
$k_y=0, k_z=\pm Q$. 

 It is instructive to compare the $\mathcal{C}''=\tau^x$ for the PDW$_\pm$ state with ${\mathcal C}'=\sigma^xs^x$ for the SDW$_z$ state at $k_z=\pm Q$. For the SDW$_z$ state, the $\mathcal{C}'$ derives from the $\mathcal{C}$-symmetry in the metallic state. For the PDW$_\pm$ state, however, the $\mathcal{C}''$ comes instead from the superconducting Nambu space, just like the case for a conventional superconductor. 
Note also that, $\mathcal{C}''$ is technically a redundancy, just like that for the BdG Hamiltonian of a topological superconductor. Therefore, the zero modes at $k_y=0, k_z=\pm Q$ only carry half of the fermionic degree of freedom, and are Majorana modes.

\section{ Drumhead surface states of the CDW  bulk}\label{section:cdw}
\subsection{Main results}
 Although the CDW instability is not the leading one, it may emerge by modifying some parameters, or simply introduced extrinsically. For example, a CDW order can be imposed either by replacing lattice atoms or by changing  bond lengths periodically.
 In the CDW state, by projecting the nodal line (see Table \ref{table:node}) onto the $xy$-surfaces,  
 there exist surface states with energies inside the bulk CDW gap~\cite{Chan:2015aa,bian_2016,Matsuura2013,Ramamurthy:2015aa}. These so-called ``drumhead" surface states only occupy a portion [the cyan regions in Fig.~\ref{fig2}(b)] of the surface BZ.  The drumhead states localized on opposite surfaces become degenerate if an additional $\mathcal {TI}$ symmetry is  imposed~\cite{Ramamurthy:2015aa}. Due to this degeneracy,  there exists a topological charge polarization between the two $xy$-surfaces. The charge density on each surface is given~\cite{Ramamurthy:2015aa} by 
 $j^0=\pm e/2 \times{b_0^2}/{4\pi}$,
  where $\pi b_0^2$ is the area enclosed by the line node projection. For a period-two CDW order ($2Q=\pi$) with even number of sites, we identify the microscopic form of  this $\mathcal{TI}$ symmetry and show that it is present when the CDW is a bond order, leading to a periodic modulation in the nearest-neighbor hopping matrix. {This $\mathcal{TI}$ symmetry also protects the CDW line node.}
   This can be experimentally realized, e.g.,  by modulating the thicknesses of the TI-SI layers~\cite{burkov_balents2011,burkov2012}. On the other hand, 
 surface states exist within line sections by projecting the bulk nodal lines onto the  $xz$- and $yz$-surfaces, which coexist and connect with the Fermi arcs [see Fig.~\ref{fig2}(b)]. 
 There are also similar drumhead surface states for both SDW$_{x}$ and SDW$_{y}$ states.  However,   the $\mathcal{TI}$ symmetry that protects the degeneracy between the two surface bands is absent.

 \subsection{Details of analysis}
 
 In the following, we analyze the $xy$ surface states for a CDW ordered bulk. The CDW order parameter, when projected to FS$_\pm$, has a projective form factor $\chi_\rho=\cos \theta$ with a line node at $\theta=\pi/2$. For definiteness we consider a special case when $2Q=\pi$, i.e., the period of the CDW is two. The analysis can be accordingly generalized to more general values of $Q$.

It is known that line nodes in the bulk projects onto the surface with in-gap surface states that exist within the momentum range bounded by the line node~\cite{Chan:2015aa,bian_2016,Matsuura2013,Ramamurthy:2015aa}.  The existence of the surface states from the line node can be understood as follows~\cite{Matsuura2013,Ramamurthy:2015aa}. We can consider a Wilson loop around the line node shown in Fig.\ \ref{sup:linenode}(a). In the 2D plane of the loop, the loop traps a Dirac node, which carries a Berry flux of $\pi$. As a result, along this loop the fermion acquires a Berry phase of $\pi$. One can distort the contour into two paths [shown in Fig.\ \ref{sup:linenode}(b)], and the berry phase difference along the two paths is given by $P_a-P_b=\pi$. We can view the two paths alternatively as along two gapped 1D subsystems $a$ and $b$, and the quantity $P_{a,b}/(2\pi)$ is the charge polarization of each 1D insulator. Therefore, the polarization of $a$ and $b$ differs by $1/2$. On the boundary this indicates that there exist edge states either inside or outside the projection of the line node. For our case, since the topology is due to the Weyl physics at small momentum $k_{x,y}$, the edge states are located inside the line node projection.

On the other hand, if the system additionally possesses a composite symmetry of an antiunitary operation ($\mathcal T$) and spatial inversion ($\mathcal I$), the polarization of all 1D subsystems are quantized to 0 or $1/2$ up to integers~\cite{Ramamurthy:2015aa}. 
{In this case the polarization for different 1D subsystems can only change discontinuously across the line node. This also indicates that the line node cannot be gapped out in the presence of $\mathcal{TI}$, since otherwise The Berry curvature would be smeared around the line node and the 1D polarizations varies smoothly~\cite{Ramamurthy:2015aa} }
For the 3D system, the total surface charge polarization can be obtained by summing the charge polarization of the topological 1D subsystems~\cite{Ramamurthy:2015aa}
\begin{align}
j^0=\pm\frac{e}{2}\frac{S_{xy}}{4\pi^2}=\pm\frac{eb_0^2}{8\pi},
\end{align}
where we have used the fact that the area enclosed by the line node $S_{xy}=\pi b_0^2$. 

{ Below we specify what the $\mathcal{TI}$ symmetry is} for the period-two CDW state of the $\mathcal{C}$-WM with wavevector $2Q=\pi$. For a translationally invariant system, the inversion operation simply corresponds to $\bm k\to -\bm k$. However, for a density-wave state, the lattice translation symmetry is broken and the inversion operation is more subtle. In this case it is easier to work in the real space rather than in $\bm k$-space.

\begin{figure}[h]
\includegraphics[width=0.5\columnwidth]{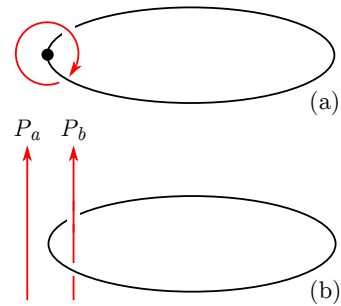}
\caption{The line node in the CDW state. The CDW wavevector is along the vertical direction, which folds the momentum space.}
\label{sup:linenode}
\end{figure}

For a given $k_x$ and $k_y$,   the Hamiltonian of the CDW ordered $\mathcal{C}$-WM  can be regarded as a 1D model  along the $z$ direction with open boundary condition:
\begin{align}
H^{k_x,k_y}_\rho=&[\sin {k_x}\sigma^x+\sin k_y\sigma^y\nonumber\\
&+(2-\cos k_x-\cos k_y)\sigma^z]\otimes\mathbb{I}_N\nonumber\\
&+\sigma^z \otimes\mathbb{C}_N +\sigma^0\otimes\mathbb{S}_N+\sigma^0\otimes \rho_N
\label{sup:hr}
\end{align}
where $\mathbb{I}_N,\mathbb{C}_N,\mathbb{S}_N,\mathbb{\rho}_N$ are matrices in coordinate space along $z$-direction with site number $N$. $\mathbb{I}_N$ is an identity matrix, and $\mathbb{C}_N,\mathbb{S}_N$ are real and imaginary nearest-neighbor hopping matrices with the form (examples are given for $N=8$)
\begin{widetext}
\begin{align}
&\mathbb{C}_N=\frac12\(\begin{array}{cccccccc} 0& 1& 0&0& 0& 0&0&0 \\1& 0& 1&0& 0& 0&0&0 \\0& 1& 0&1& 0& 0&0 &0\\0& 0& 1&0& 1& 0&0&0 \\0& 0& 0&1& 0& 1&0 &0\\0& 0& 0&0& 1& 0&1&0 \\0& 0& 0&0& 0& 1&0 &1\\0& 0& 0&0& 0&0& 1&0 \end{array}\),~
\mathbb{S}_N=-\frac{i}2\(\begin{array}{cccccccc} 0& 1& 0&0& 0& 0&0&0 \\-1& 0& 1&0& 0& 0&0&0 \\0& -1& 0&1& 0& 0&0 &0\\0& 0& -1&0& 1& 0&0&0 \\0& 0& 0&-1& 0& 1&0 &0\\0& 0& 0&0& -1& 0&1&0 \\0& 0& 0&0& 0& -1&0 &1\\0& 0& 0&0& 0&0& -1&0\end{array}\),\nonumber
\end{align}
\end{widetext}

and $\rho_N$ is the CDW matrix which we will address.
One can verify that the first three terms in Eq.\ \eqref{sup:hr} with $\mathbb{I}_N,\mathbb{C}_N,\mathbb{S}_N$ preserve a composite symmetry of $\mathcal{TI}$, where 
\begin{align}
\mathcal{T}= &\exp \[-i\frac{\sigma^z}{2} \arctan\frac{\sin k_y}{\sin k_x} \]K\exp \[i\frac{\sigma^z}{2} \arctan\frac{\sin k_y}{\sin k_x} \] \nonumber\\
=&\exp \[-i{\sigma^z} \arctan\frac{\sin k_y}{\sin k_x} \]K
\label{sup:T}
\end{align}
 is an anti-unitary operator  ($K$ is the usual complex conjugation sending $i$ to $-i$), and $\mathcal{I}$ is spatial inversion that takes $z\to N+1-z$, namely,
 \begin{align}
 \mathcal{I}=\(\begin{array}{cccccccc} 0& 0& 0&0& 0& 0&0&1 \\0& 0& 0&0& 0& 0&1&0 \\0& 0& 0&0& 0& 1&0 &0\\0& 0& 0&0& 1& 0&0&0 \\0& 0& 0&1& 0& 0&0 &0\\0& 0& 1&0& 0& 0&0&0 \\0& 1& 0&0& 0& 0&0 &0\\1& 0& 0&0& 0&0& 0&0\end{array}\).
 \label{sup:I}
  \end{align}
  For the CDW term with wavevector $2Q=\pi$, the unit cell size is doubled, and we specifically focus on a case with even number of sites $N=2m$. We consider two types of the CDW's of the form,
\begin{align}
\mathbb{\rho}^{(a)}_N=\frac12&\(\begin{array}{cccccccc} 1& 0& 0&0& 0& 0&0&0 \\0& -1& 0&0& 0& 0&0&0 \\0& 0& 1&0& 0& 0&0 &0\\0& 0& 0&-1& 0& 0&0&0 \\0& 0& 0&0& 1& 0&0 &0\\0& 0& 0&0& 0& -1&0&0 \\0& 0& 0&0& 0& 0&1 &0\\0& 0& 0&0& 0& 0&0 &-1\end{array}\),\nonumber\\
\mathbb{\rho}^{(b)}_N=-\frac{1}2&\(\begin{array}{cccccccc} 0& 1& 0&0& 0& 0&0&0 \\1& 0& -1&0& 0& 0&0&0 \\0& -1& 0&1& 0& 0&0 &0\\0& 0& 1&0& -1& 0&0&0 \\0& 0& 0&-1& 0& 1&0 &0\\0& 0& 0&0& 1& 0&-1&0 \\0& 0& 0&0& 0& -1&0 &1\\0&0& 0& 0&0& 0& 1&0 \end{array}\),\nonumber
\end{align}
where the first one is a site order that modulates the on-site charge density and the second one is a bond order that modulates the nearest-neighbor hopping amplitudes. In $\bm k$ space, the two types of CDW's are related by a $\pi/2$ phase shift of the order parameter.
 It is easy to see that both are invariant under $\mathcal T$, however only $\rho_N^{(b)}$, the bond order, is invariant under $\mathcal{I}$ (which takes $z\to N+1-z$), while $\rho_N^{(a)}$, the site order, reverses sign under $\mathcal{I}$. On the other hand it is straightforward to see that for \emph{odd} number of sites, the $\mathcal{TI}$ symmetry is present for a CDW site order. This even-odd dichotomy is similar to case of edge states in graphene~\cite{bernevig_book}.
 
\begin{figure*}
\begin{tabular}{cc}
\includegraphics[width=0.8\columnwidth]{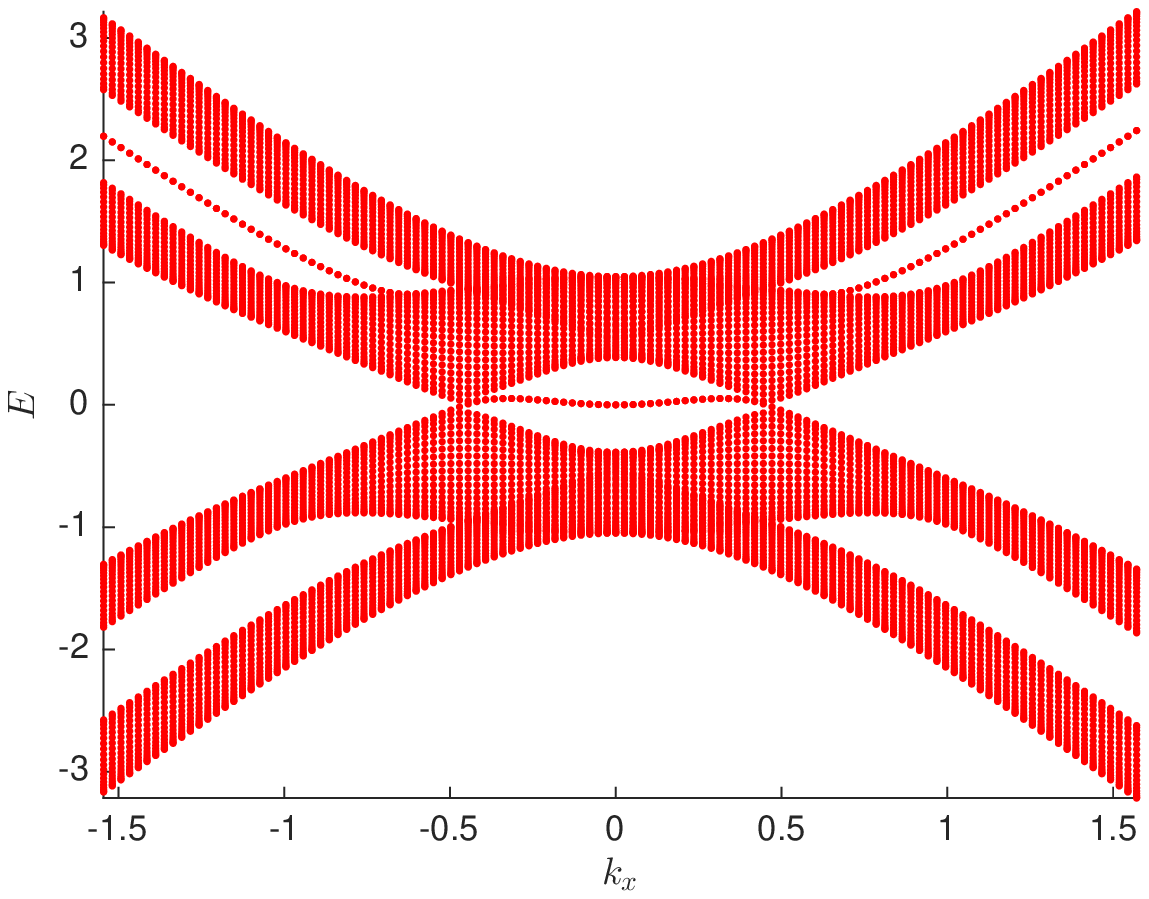}&\includegraphics[width=0.8\columnwidth]{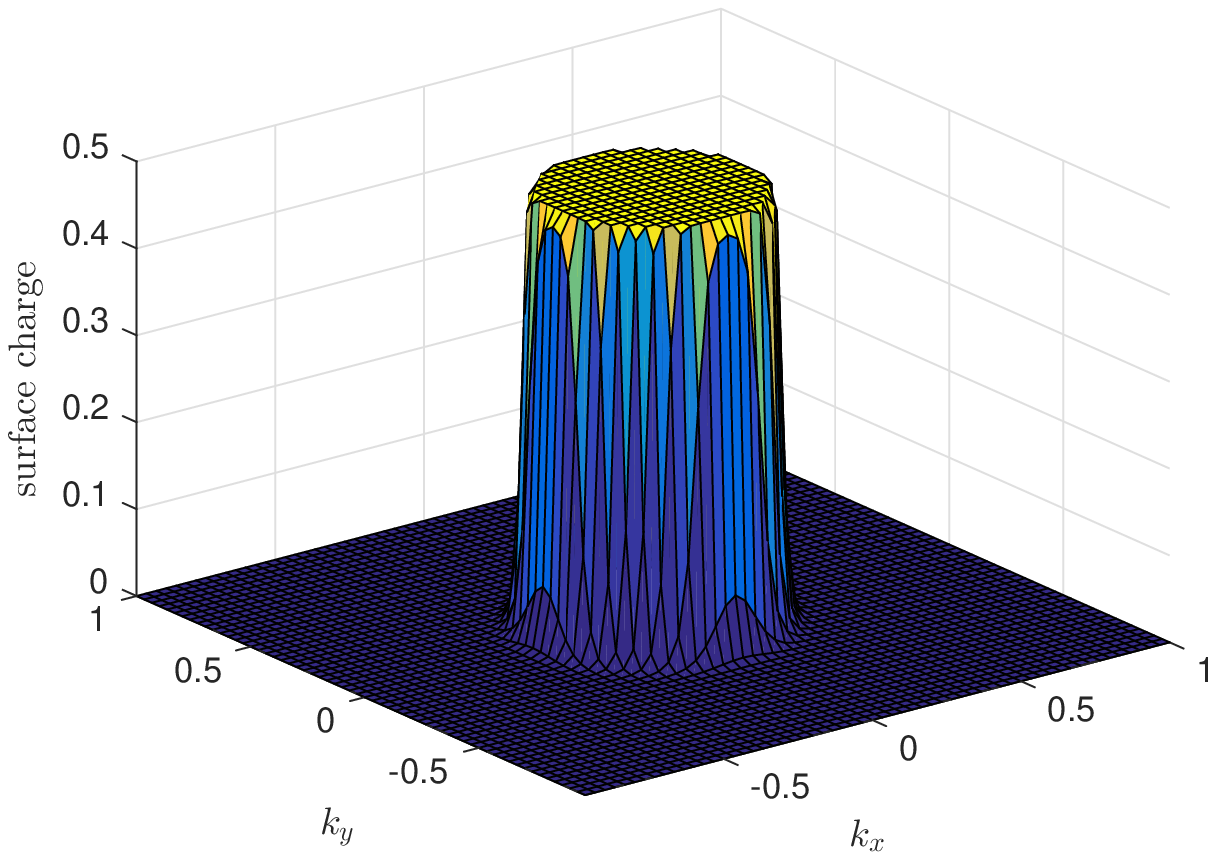}\\
(a)&(b)
\end{tabular}
\caption{(a) The in-gap drumhead bands ($|k_x|\lesssim0.5$)  for a bond-ordered CDW state at $k_y=0$ with open boundaries in $z$-direction. Each in-gap point is two-fold degenerate corresponding to two opposite surface states. (b) The surface charge density in the $xy$ surface BZ due to the line node in the bulk.}
\label{sup:drumhead}
\end{figure*}

 With the $\mathcal{TI}$ symmetry defined by Eqs.\ (\ref{sup:T}) and (\ref{sup:I}), it is straightforward to show that the charge polarization along $z$ direction for 1D subsystems with any given $k_{x,y}$ is quantized. The charge polarization along $z$ satisfies
 \begin{align}
 P_z(k_x, k_y)=&\frac{e}{2\pi}\int_{-\pi}^{\pi} dk_z \sum_{E_a(\bm k)<0}\langle u_a(\bm k)| z|u_a(\bm k)\rangle \nonumber\\
 =&\frac{e}{2\pi}\int_{-\pi}^{\pi} dk_z \sum_{E_a(\bm k)<0} \langle u_a(\bm k)|(\mathcal{TI}) z(\mathcal{TI})|u_a(\bm k)\rangle \nonumber\\
 =&-\frac{e}{2\pi}\int_{-\pi}^{\pi} dk_z \sum_{E_a'(\bm k)<0} \langle u_a'(\bm k)|\hat z|u_a'(\bm k)\rangle  \nonumber\\
 =&-P_z(k_x, k_y).
 \label{sup:pz}
 \end{align}
 Since $P_z$ is well-defined only up to an integer, its value is quantized to either $0$ or $1/2$.  
 {As we discussed, this also indicates the nodal line in the bulk is protected.}
  
We numerically computed the energy spectrum for $\rho_N^{(b)}$ bond order with open boundary conditions in $z$ direction with $N=40$, $2Q=\pi$, $k_y=0$, $b_0= 0.3$ and $\rho=0.2$. We show the resulting band structure and surface charge density in Fig.\ \ref{sup:drumhead}.

\section{Conclusion}\label{section:conclusion}
 
 In this paper, we  studied the instabilities of a particle-hole symmetric Weyl metal with both repulsive and attractive interactions. We analyzed the  nodal structures of each order (Table \ref{table:node}) and addressed their relation to the Berry curvature flux through the FS's. As a result of distinct nodal structures, we found that the leading instabilities are SDW$_z$ (for a repulsive interaction) and PDW$_\pm$ orders (for an attractive interaction), with wavevectors equal to the separation between the Weyl points. We analyzed the properties of surface states, schematically shown in Fig.~\ref{fig2}. We found that in the SDW$_z$ and PDW states, the surface Fermi arc in the $\mathcal{C}$-WM state now goes through the projection of the Weyl points and traverses the full BZ.
In the CDW state, there exist drumhead surface bands inside the projection of the nodal line, and can lead to topologically protected charge responses on the surface. 
Experimentally, the SDW state realizes a 3D Hall effect, and  can be detected via neutron scattering or nulcear magnetic resonance measurements. The PDW states can be detected via a scanning Josephson tunneling microscopy that has been recently developed~\cite{davis_last}. The surface states can be detected using ARPES.  It will be interesting to search for these orders in Weyl materials in the near future. 

\acknowledgments We thank J. G. Analytis, G. Y. Cho, A. V. Chubukov,  R. Fernandes, E. Fradkin, T. L. Hughes, J. Kang, S. Ramamurthy and F. Yang for inspiring discussions. This work was supported in part by the Gordon and Betty Moore Foundation's EPiQS Initiative through Grant No. GBMF4305 at the University of Illinois (Y.W.) and by the NSF through grant No. DMR 1408713 at the University of Illinois (P.Y.).


\begin{thebibliography}{71}%
\makeatletter
\providecommand \@ifxundefined [1]{%
 \@ifx{#1\undefined}
}%
\providecommand \@ifnum [1]{%
 \ifnum #1\expandafter \@firstoftwo
 \else \expandafter \@secondoftwo
 \fi
}%
\providecommand \@ifx [1]{%
 \ifx #1\expandafter \@firstoftwo
 \else \expandafter \@secondoftwo
 \fi
}%
\providecommand \natexlab [1]{#1}%
\providecommand \enquote  [1]{``#1''}%
\providecommand \bibnamefont  [1]{#1}%
\providecommand \bibfnamefont [1]{#1}%
\providecommand \citenamefont [1]{#1}%
\providecommand \href@noop [0]{\@secondoftwo}%
\providecommand \href [0]{\begingroup \@sanitize@url \@href}%
\providecommand \@href[1]{\@@startlink{#1}\@@href}%
\providecommand \@@href[1]{\endgroup#1\@@endlink}%
\providecommand \@sanitize@url [0]{\catcode `\\12\catcode `\$12\catcode
  `\&12\catcode `\#12\catcode `\^12\catcode `\_12\catcode `\%12\relax}%
\providecommand \@@startlink[1]{}%
\providecommand \@@endlink[0]{}%
\providecommand \url  [0]{\begingroup\@sanitize@url \@url }%
\providecommand \@url [1]{\endgroup\@href {#1}{\urlprefix }}%
\providecommand \urlprefix  [0]{URL }%
\providecommand \Eprint [0]{\href }%
\providecommand \doibase [0]{http://dx.doi.org/}%
\providecommand \selectlanguage [0]{\@gobble}%
\providecommand \bibinfo  [0]{\@secondoftwo}%
\providecommand \bibfield  [0]{\@secondoftwo}%
\providecommand \translation [1]{[#1]}%
\providecommand \BibitemOpen [0]{}%
\providecommand \bibitemStop [0]{}%
\providecommand \bibitemNoStop [0]{.\EOS\space}%
\providecommand \EOS [0]{\spacefactor3000\relax}%
\providecommand \BibitemShut  [1]{\csname bibitem#1\endcsname}%
\let\auto@bib@innerbib\@empty
\bibitem [{\citenamefont {Wan}\ \emph {et~al.}(2011)\citenamefont {Wan},
  \citenamefont {Turner}, \citenamefont {Vishwanath},\ and\ \citenamefont
  {Savrasov}}]{xwan11}%
  \BibitemOpen
  \bibfield  {author} {\bibinfo {author} {\bibfnamefont {X.}~\bibnamefont
  {Wan}}, \bibinfo {author} {\bibfnamefont {A.~M.}\ \bibnamefont {Turner}},
  \bibinfo {author} {\bibfnamefont {A.}~\bibnamefont {Vishwanath}}, \ and\
  \bibinfo {author} {\bibfnamefont {S.~Y.}\ \bibnamefont {Savrasov}},\ }\href
  {\doibase 10.1103/PhysRevB.83.205101} {\bibfield  {journal} {\bibinfo
  {journal} {Phys. Rev. B}\ }\textbf {\bibinfo {volume} {83}},\ \bibinfo
  {pages} {205101} (\bibinfo {year} {2011})}\BibitemShut {NoStop}%
\bibitem [{\citenamefont {Xu}\ \emph {et~al.}(2011)\citenamefont {Xu},
  \citenamefont {Weng}, \citenamefont {Wang}, \citenamefont {Dai},\ and\
  \citenamefont {Fang}}]{xugang11}%
  \BibitemOpen
  \bibfield  {author} {\bibinfo {author} {\bibfnamefont {G.}~\bibnamefont
  {Xu}}, \bibinfo {author} {\bibfnamefont {H.}~\bibnamefont {Weng}}, \bibinfo
  {author} {\bibfnamefont {Z.}~\bibnamefont {Wang}}, \bibinfo {author}
  {\bibfnamefont {X.}~\bibnamefont {Dai}}, \ and\ \bibinfo {author}
  {\bibfnamefont {Z.}~\bibnamefont {Fang}},\ }\href {\doibase
  10.1103/PhysRevLett.107.186806} {\bibfield  {journal} {\bibinfo  {journal}
  {Phys. Rev. Lett.}\ }\textbf {\bibinfo {volume} {107}},\ \bibinfo {pages}
  {186806} (\bibinfo {year} {2011})}\BibitemShut {NoStop}%
\bibitem [{\citenamefont {Burkov}\ and\ \citenamefont
  {Balents}(2011)}]{burkov_balents2011}%
  \BibitemOpen
  \bibfield  {author} {\bibinfo {author} {\bibfnamefont {A.~A.}\ \bibnamefont
  {Burkov}}\ and\ \bibinfo {author} {\bibfnamefont {L.}~\bibnamefont
  {Balents}},\ }\href {\doibase 10.1103/PhysRevLett.107.127205} {\bibfield
  {journal} {\bibinfo  {journal} {Phys. Rev. Lett.}\ }\textbf {\bibinfo
  {volume} {107}},\ \bibinfo {pages} {127205} (\bibinfo {year}
  {2011})}\BibitemShut {NoStop}%
\bibitem [{\citenamefont {Yang}\ \emph {et~al.}(2011)\citenamefont {Yang},
  \citenamefont {Lu},\ and\ \citenamefont {Ran}}]{YLR2011}%
  \BibitemOpen
  \bibfield  {author} {\bibinfo {author} {\bibfnamefont {K.-Y.}\ \bibnamefont
  {Yang}}, \bibinfo {author} {\bibfnamefont {Y.-M.}\ \bibnamefont {Lu}}, \ and\
  \bibinfo {author} {\bibfnamefont {Y.}~\bibnamefont {Ran}},\ }\href {\doibase
  10.1103/PhysRevB.84.075129} {\bibfield  {journal} {\bibinfo  {journal} {Phys.
  Rev. B}\ }\textbf {\bibinfo {volume} {84}},\ \bibinfo {pages} {075129}
  (\bibinfo {year} {2011})}\BibitemShut {NoStop}%
\bibitem [{\citenamefont {Xu}\ \emph {et~al.}(2015{\natexlab{a}})\citenamefont
  {Xu}, \citenamefont {Belopolski}, \citenamefont {Alidoust}, \citenamefont
  {Neupane}, \citenamefont {Bian}, \citenamefont {Zhang}, \citenamefont
  {Sankar}, \citenamefont {Chang}, \citenamefont {Yuan}, \citenamefont {Lee},
  \citenamefont {Huang}, \citenamefont {Zheng}, \citenamefont {Ma},
  \citenamefont {Sanchez}, \citenamefont {Wang}, \citenamefont {Bansil},
  \citenamefont {Chou}, \citenamefont {Shibayev}, \citenamefont {Lin},
  \citenamefont {Jia},\ and\ \citenamefont {Hasan}}]{Xu613}%
  \BibitemOpen
  \bibfield  {author} {\bibinfo {author} {\bibfnamefont {S.-Y.}\ \bibnamefont
  {Xu}}, \bibinfo {author} {\bibfnamefont {I.}~\bibnamefont {Belopolski}},
  \bibinfo {author} {\bibfnamefont {N.}~\bibnamefont {Alidoust}}, \bibinfo
  {author} {\bibfnamefont {M.}~\bibnamefont {Neupane}}, \bibinfo {author}
  {\bibfnamefont {G.}~\bibnamefont {Bian}}, \bibinfo {author} {\bibfnamefont
  {C.}~\bibnamefont {Zhang}}, \bibinfo {author} {\bibfnamefont
  {R.}~\bibnamefont {Sankar}}, \bibinfo {author} {\bibfnamefont
  {G.}~\bibnamefont {Chang}}, \bibinfo {author} {\bibfnamefont
  {Z.}~\bibnamefont {Yuan}}, \bibinfo {author} {\bibfnamefont {C.-C.}\
  \bibnamefont {Lee}}, \bibinfo {author} {\bibfnamefont {S.-M.}\ \bibnamefont
  {Huang}}, \bibinfo {author} {\bibfnamefont {H.}~\bibnamefont {Zheng}},
  \bibinfo {author} {\bibfnamefont {J.}~\bibnamefont {Ma}}, \bibinfo {author}
  {\bibfnamefont {D.~S.}\ \bibnamefont {Sanchez}}, \bibinfo {author}
  {\bibfnamefont {B.}~\bibnamefont {Wang}}, \bibinfo {author} {\bibfnamefont
  {A.}~\bibnamefont {Bansil}}, \bibinfo {author} {\bibfnamefont
  {F.}~\bibnamefont {Chou}}, \bibinfo {author} {\bibfnamefont {P.~P.}\
  \bibnamefont {Shibayev}}, \bibinfo {author} {\bibfnamefont {H.}~\bibnamefont
  {Lin}}, \bibinfo {author} {\bibfnamefont {S.}~\bibnamefont {Jia}}, \ and\
  \bibinfo {author} {\bibfnamefont {M.~Z.}\ \bibnamefont {Hasan}},\ }\href
  {\doibase 10.1126/science.aaa9297} {\bibfield  {journal} {\bibinfo  {journal}
  {Science}\ }\textbf {\bibinfo {volume} {349}},\ \bibinfo {pages} {613}
  (\bibinfo {year} {2015}{\natexlab{a}})}\BibitemShut {NoStop}%
\bibitem [{\citenamefont {Lv}\ \emph {et~al.}(2015{\natexlab{a}})\citenamefont
  {Lv}, \citenamefont {Weng}, \citenamefont {Fu}, \citenamefont {Wang},
  \citenamefont {Miao}, \citenamefont {Ma}, \citenamefont {Richard},
  \citenamefont {Huang}, \citenamefont {Zhao}, \citenamefont {Chen},
  \citenamefont {Fang}, \citenamefont {Dai}, \citenamefont {Qian},\ and\
  \citenamefont {Ding}}]{Lv2015}%
  \BibitemOpen
  \bibfield  {author} {\bibinfo {author} {\bibfnamefont {B.~Q.}\ \bibnamefont
  {Lv}}, \bibinfo {author} {\bibfnamefont {H.~M.}\ \bibnamefont {Weng}},
  \bibinfo {author} {\bibfnamefont {B.~B.}\ \bibnamefont {Fu}}, \bibinfo
  {author} {\bibfnamefont {X.~P.}\ \bibnamefont {Wang}}, \bibinfo {author}
  {\bibfnamefont {H.}~\bibnamefont {Miao}}, \bibinfo {author} {\bibfnamefont
  {J.}~\bibnamefont {Ma}}, \bibinfo {author} {\bibfnamefont {P.}~\bibnamefont
  {Richard}}, \bibinfo {author} {\bibfnamefont {X.~C.}\ \bibnamefont {Huang}},
  \bibinfo {author} {\bibfnamefont {L.~X.}\ \bibnamefont {Zhao}}, \bibinfo
  {author} {\bibfnamefont {G.~F.}\ \bibnamefont {Chen}}, \bibinfo {author}
  {\bibfnamefont {Z.}~\bibnamefont {Fang}}, \bibinfo {author} {\bibfnamefont
  {X.}~\bibnamefont {Dai}}, \bibinfo {author} {\bibfnamefont {T.}~\bibnamefont
  {Qian}}, \ and\ \bibinfo {author} {\bibfnamefont {H.}~\bibnamefont {Ding}},\
  }\href {\doibase 10.1103/PhysRevX.5.031013} {\bibfield  {journal} {\bibinfo
  {journal} {Phys. Rev. X}\ }\textbf {\bibinfo {volume} {5}},\ \bibinfo {pages}
  {031013} (\bibinfo {year} {2015}{\natexlab{a}})}\BibitemShut {NoStop}%
\bibitem [{\citenamefont {Xu}\ \emph {et~al.}(2015{\natexlab{b}})\citenamefont
  {Xu}, \citenamefont {Alidoust}, \citenamefont {Belopolski}, \citenamefont
  {Yuan}, \citenamefont {Bian}, \citenamefont {Chang}, \citenamefont {Zheng},
  \citenamefont {Strocov}, \citenamefont {Sanchez}, \citenamefont {Chang},
  \citenamefont {Zhang}, \citenamefont {Mou}, \citenamefont {Wu}, \citenamefont
  {Huang}, \citenamefont {Lee}, \citenamefont {Huang}, \citenamefont {Wang},
  \citenamefont {Bansil}, \citenamefont {Jeng}, \citenamefont {Neupert},
  \citenamefont {Kaminski}, \citenamefont {Lin}, \citenamefont {Jia},\ and\
  \citenamefont {Zahid~Hasan}}]{Xu:2015aa}%
  \BibitemOpen
  \bibfield  {author} {\bibinfo {author} {\bibfnamefont {S.-Y.}\ \bibnamefont
  {Xu}}, \bibinfo {author} {\bibfnamefont {N.}~\bibnamefont {Alidoust}},
  \bibinfo {author} {\bibfnamefont {I.}~\bibnamefont {Belopolski}}, \bibinfo
  {author} {\bibfnamefont {Z.}~\bibnamefont {Yuan}}, \bibinfo {author}
  {\bibfnamefont {G.}~\bibnamefont {Bian}}, \bibinfo {author} {\bibfnamefont
  {T.-R.}\ \bibnamefont {Chang}}, \bibinfo {author} {\bibfnamefont
  {H.}~\bibnamefont {Zheng}}, \bibinfo {author} {\bibfnamefont {V.~N.}\
  \bibnamefont {Strocov}}, \bibinfo {author} {\bibfnamefont {D.~S.}\
  \bibnamefont {Sanchez}}, \bibinfo {author} {\bibfnamefont {G.}~\bibnamefont
  {Chang}}, \bibinfo {author} {\bibfnamefont {C.}~\bibnamefont {Zhang}},
  \bibinfo {author} {\bibfnamefont {D.}~\bibnamefont {Mou}}, \bibinfo {author}
  {\bibfnamefont {Y.}~\bibnamefont {Wu}}, \bibinfo {author} {\bibfnamefont
  {L.}~\bibnamefont {Huang}}, \bibinfo {author} {\bibfnamefont {C.-C.}\
  \bibnamefont {Lee}}, \bibinfo {author} {\bibfnamefont {S.-M.}\ \bibnamefont
  {Huang}}, \bibinfo {author} {\bibfnamefont {B.}~\bibnamefont {Wang}},
  \bibinfo {author} {\bibfnamefont {A.}~\bibnamefont {Bansil}}, \bibinfo
  {author} {\bibfnamefont {H.-T.}\ \bibnamefont {Jeng}}, \bibinfo {author}
  {\bibfnamefont {T.}~\bibnamefont {Neupert}}, \bibinfo {author} {\bibfnamefont
  {A.}~\bibnamefont {Kaminski}}, \bibinfo {author} {\bibfnamefont
  {H.}~\bibnamefont {Lin}}, \bibinfo {author} {\bibfnamefont {S.}~\bibnamefont
  {Jia}}, \ and\ \bibinfo {author} {\bibfnamefont {M.}~\bibnamefont
  {Zahid~Hasan}},\ }\href {http://dx.doi.org/10.1038/nphys3437} {\bibfield
  {journal} {\bibinfo  {journal} {Nat. Phys.}\ }\textbf {\bibinfo {volume}
  {11}},\ \bibinfo {pages} {748} (\bibinfo {year}
  {2015}{\natexlab{b}})}\BibitemShut {NoStop}%
\bibitem [{\citenamefont {Lv}\ \emph {et~al.}(2015{\natexlab{b}})\citenamefont
  {Lv}, \citenamefont {Xu}, \citenamefont {Weng}, \citenamefont {Ma},
  \citenamefont {Richard}, \citenamefont {Huang}, \citenamefont {Zhao},
  \citenamefont {Chen}, \citenamefont {Matt}, \citenamefont {Bisti},
  \citenamefont {Strocov}, \citenamefont {Mesot}, \citenamefont {Fang},
  \citenamefont {Dai}, \citenamefont {Qian}, \citenamefont {Shi},\ and\
  \citenamefont {Ding}}]{Lv:2015aa}%
  \BibitemOpen
  \bibfield  {author} {\bibinfo {author} {\bibfnamefont {B.~Q.}\ \bibnamefont
  {Lv}}, \bibinfo {author} {\bibfnamefont {N.}~\bibnamefont {Xu}}, \bibinfo
  {author} {\bibfnamefont {H.~M.}\ \bibnamefont {Weng}}, \bibinfo {author}
  {\bibfnamefont {J.~Z.}\ \bibnamefont {Ma}}, \bibinfo {author} {\bibfnamefont
  {P.}~\bibnamefont {Richard}}, \bibinfo {author} {\bibfnamefont {X.~C.}\
  \bibnamefont {Huang}}, \bibinfo {author} {\bibfnamefont {L.~X.}\ \bibnamefont
  {Zhao}}, \bibinfo {author} {\bibfnamefont {G.~F.}\ \bibnamefont {Chen}},
  \bibinfo {author} {\bibfnamefont {C.~E.}\ \bibnamefont {Matt}}, \bibinfo
  {author} {\bibfnamefont {F.}~\bibnamefont {Bisti}}, \bibinfo {author}
  {\bibfnamefont {V.~N.}\ \bibnamefont {Strocov}}, \bibinfo {author}
  {\bibfnamefont {J.}~\bibnamefont {Mesot}}, \bibinfo {author} {\bibfnamefont
  {Z.}~\bibnamefont {Fang}}, \bibinfo {author} {\bibfnamefont {X.}~\bibnamefont
  {Dai}}, \bibinfo {author} {\bibfnamefont {T.}~\bibnamefont {Qian}}, \bibinfo
  {author} {\bibfnamefont {M.}~\bibnamefont {Shi}}, \ and\ \bibinfo {author}
  {\bibfnamefont {H.}~\bibnamefont {Ding}},\ }\href
  {http://dx.doi.org/10.1038/nphys3426} {\bibfield  {journal} {\bibinfo
  {journal} {Nat. Phys.}\ }\textbf {\bibinfo {volume} {11}},\ \bibinfo {pages}
  {724} (\bibinfo {year} {2015}{\natexlab{b}})}\BibitemShut {NoStop}%
\bibitem [{\citenamefont {Yang}\ \emph {et~al.}(2015)\citenamefont {Yang},
  \citenamefont {Liu}, \citenamefont {Sun}, \citenamefont {Peng}, \citenamefont
  {Yang}, \citenamefont {Zhang}, \citenamefont {Zhou}, \citenamefont {Zhang},
  \citenamefont {Guo}, \citenamefont {Rahn}, \citenamefont {Prabhakaran},
  \citenamefont {Hussain}, \citenamefont {Mo}, \citenamefont {Felser},
  \citenamefont {Yan},\ and\ \citenamefont {Chen}}]{Yang:2015aa}%
  \BibitemOpen
  \bibfield  {author} {\bibinfo {author} {\bibfnamefont {L.~X.}\ \bibnamefont
  {Yang}}, \bibinfo {author} {\bibfnamefont {Z.~K.}\ \bibnamefont {Liu}},
  \bibinfo {author} {\bibfnamefont {Y.}~\bibnamefont {Sun}}, \bibinfo {author}
  {\bibfnamefont {H.}~\bibnamefont {Peng}}, \bibinfo {author} {\bibfnamefont
  {H.~F.}\ \bibnamefont {Yang}}, \bibinfo {author} {\bibfnamefont
  {T.}~\bibnamefont {Zhang}}, \bibinfo {author} {\bibfnamefont
  {B.}~\bibnamefont {Zhou}}, \bibinfo {author} {\bibfnamefont {Y.}~\bibnamefont
  {Zhang}}, \bibinfo {author} {\bibfnamefont {Y.~F.}\ \bibnamefont {Guo}},
  \bibinfo {author} {\bibfnamefont {M.}~\bibnamefont {Rahn}}, \bibinfo {author}
  {\bibfnamefont {D.}~\bibnamefont {Prabhakaran}}, \bibinfo {author}
  {\bibfnamefont {Z.}~\bibnamefont {Hussain}}, \bibinfo {author} {\bibfnamefont
  {S.~K.}\ \bibnamefont {Mo}}, \bibinfo {author} {\bibfnamefont
  {C.}~\bibnamefont {Felser}}, \bibinfo {author} {\bibfnamefont
  {B.}~\bibnamefont {Yan}}, \ and\ \bibinfo {author} {\bibfnamefont {Y.~L.}\
  \bibnamefont {Chen}},\ }\href@noop {} {\bibfield  {journal} {\bibinfo
  {journal} {Nat. Phys.}\ }\textbf {\bibinfo {volume} {11}},\ \bibinfo {pages}
  {728} (\bibinfo {year} {2015})}\BibitemShut {NoStop}%
\bibitem [{\citenamefont {Xu}\ \emph {et~al.}(2015{\natexlab{c}})\citenamefont
  {Xu}, \citenamefont {Belopolski}, \citenamefont {Sanchez}, \citenamefont
  {Zhang}, \citenamefont {Chang}, \citenamefont {Guo}, \citenamefont {Bian},
  \citenamefont {Yuan}, \citenamefont {Lu}, \citenamefont {Chang},
  \citenamefont {Shibayev}, \citenamefont {Prokopovych}, \citenamefont
  {Alidoust}, \citenamefont {Zheng}, \citenamefont {Lee}, \citenamefont
  {Huang}, \citenamefont {Sankar}, \citenamefont {Chou}, \citenamefont {Hsu},
  \citenamefont {Jeng}, \citenamefont {Bansil}, \citenamefont {Neupert},
  \citenamefont {Strocov}, \citenamefont {Lin}, \citenamefont {Jia},\ and\
  \citenamefont {Hasan}}]{Xue1501092}%
  \BibitemOpen
  \bibfield  {author} {\bibinfo {author} {\bibfnamefont {S.-Y.}\ \bibnamefont
  {Xu}}, \bibinfo {author} {\bibfnamefont {I.}~\bibnamefont {Belopolski}},
  \bibinfo {author} {\bibfnamefont {D.~S.}\ \bibnamefont {Sanchez}}, \bibinfo
  {author} {\bibfnamefont {C.}~\bibnamefont {Zhang}}, \bibinfo {author}
  {\bibfnamefont {G.}~\bibnamefont {Chang}}, \bibinfo {author} {\bibfnamefont
  {C.}~\bibnamefont {Guo}}, \bibinfo {author} {\bibfnamefont {G.}~\bibnamefont
  {Bian}}, \bibinfo {author} {\bibfnamefont {Z.}~\bibnamefont {Yuan}}, \bibinfo
  {author} {\bibfnamefont {H.}~\bibnamefont {Lu}}, \bibinfo {author}
  {\bibfnamefont {T.-R.}\ \bibnamefont {Chang}}, \bibinfo {author}
  {\bibfnamefont {P.~P.}\ \bibnamefont {Shibayev}}, \bibinfo {author}
  {\bibfnamefont {M.~L.}\ \bibnamefont {Prokopovych}}, \bibinfo {author}
  {\bibfnamefont {N.}~\bibnamefont {Alidoust}}, \bibinfo {author}
  {\bibfnamefont {H.}~\bibnamefont {Zheng}}, \bibinfo {author} {\bibfnamefont
  {C.-C.}\ \bibnamefont {Lee}}, \bibinfo {author} {\bibfnamefont {S.-M.}\
  \bibnamefont {Huang}}, \bibinfo {author} {\bibfnamefont {R.}~\bibnamefont
  {Sankar}}, \bibinfo {author} {\bibfnamefont {F.}~\bibnamefont {Chou}},
  \bibinfo {author} {\bibfnamefont {C.-H.}\ \bibnamefont {Hsu}}, \bibinfo
  {author} {\bibfnamefont {H.-T.}\ \bibnamefont {Jeng}}, \bibinfo {author}
  {\bibfnamefont {A.}~\bibnamefont {Bansil}}, \bibinfo {author} {\bibfnamefont
  {T.}~\bibnamefont {Neupert}}, \bibinfo {author} {\bibfnamefont {V.~N.}\
  \bibnamefont {Strocov}}, \bibinfo {author} {\bibfnamefont {H.}~\bibnamefont
  {Lin}}, \bibinfo {author} {\bibfnamefont {S.}~\bibnamefont {Jia}}, \ and\
  \bibinfo {author} {\bibfnamefont {M.~Z.}\ \bibnamefont {Hasan}},\ }\href
  {\doibase 10.1126/sciadv.1501092} {\bibfield  {journal} {\bibinfo  {journal}
  {Sci. Adv.}\ }\textbf {\bibinfo {volume} {1}} (\bibinfo {year}
  {2015}{\natexlab{c}}),\ 10.1126/sciadv.1501092},\ \Eprint
  {http://arxiv.org/abs/arXiv:1508.03102} {arXiv:1508.03102} \BibitemShut
  {NoStop}%
\bibitem [{\citenamefont {Liu}\ \emph {et~al.}(2016)\citenamefont {Liu},
  \citenamefont {Yang}, \citenamefont {Sun}, \citenamefont {Zhang},
  \citenamefont {Peng}, \citenamefont {Yang}, \citenamefont {Chen},
  \citenamefont {Zhang}, \citenamefont {Guo}, \citenamefont {Prabhakaran},
  \citenamefont {Schmidt}, \citenamefont {Hussain}, \citenamefont {Mo},
  \citenamefont {Felser}, \citenamefont {Yan},\ and\ \citenamefont
  {Chen}}]{Liu:2016aa}%
  \BibitemOpen
  \bibfield  {author} {\bibinfo {author} {\bibfnamefont {Z.~K.}\ \bibnamefont
  {Liu}}, \bibinfo {author} {\bibfnamefont {L.~X.}\ \bibnamefont {Yang}},
  \bibinfo {author} {\bibfnamefont {Y.}~\bibnamefont {Sun}}, \bibinfo {author}
  {\bibfnamefont {T.}~\bibnamefont {Zhang}}, \bibinfo {author} {\bibfnamefont
  {H.}~\bibnamefont {Peng}}, \bibinfo {author} {\bibfnamefont {H.~F.}\
  \bibnamefont {Yang}}, \bibinfo {author} {\bibfnamefont {C.}~\bibnamefont
  {Chen}}, \bibinfo {author} {\bibfnamefont {Y.}~\bibnamefont {Zhang}},
  \bibinfo {author} {\bibfnamefont {Y.~F.}\ \bibnamefont {Guo}}, \bibinfo
  {author} {\bibfnamefont {D.}~\bibnamefont {Prabhakaran}}, \bibinfo {author}
  {\bibfnamefont {M.}~\bibnamefont {Schmidt}}, \bibinfo {author} {\bibfnamefont
  {Z.}~\bibnamefont {Hussain}}, \bibinfo {author} {\bibfnamefont {S.~K.}\
  \bibnamefont {Mo}}, \bibinfo {author} {\bibfnamefont {C.}~\bibnamefont
  {Felser}}, \bibinfo {author} {\bibfnamefont {B.}~\bibnamefont {Yan}}, \ and\
  \bibinfo {author} {\bibfnamefont {Y.~L.}\ \bibnamefont {Chen}},\ }\href
  {http://dx.doi.org/10.1038/nmat4457} {\bibfield  {journal} {\bibinfo
  {journal} {Nat. Mater.}\ }\textbf {\bibinfo {volume} {15}},\ \bibinfo {pages}
  {27} (\bibinfo {year} {2016})}\BibitemShut {NoStop}%
\bibitem [{\citenamefont {Belopolski}\ \emph {et~al.}(2016)\citenamefont
  {Belopolski}, \citenamefont {Xu}, \citenamefont {Sanchez}, \citenamefont
  {Chang}, \citenamefont {Guo}, \citenamefont {Neupane}, \citenamefont {Zheng},
  \citenamefont {Lee}, \citenamefont {Huang}, \citenamefont {Bian},
  \citenamefont {Alidoust}, \citenamefont {Chang}, \citenamefont {Wang},
  \citenamefont {Zhang}, \citenamefont {Bansil}, \citenamefont {Jeng},
  \citenamefont {Lin}, \citenamefont {Jia},\ and\ \citenamefont
  {Hasan}}]{Belopolski2016}%
  \BibitemOpen
  \bibfield  {author} {\bibinfo {author} {\bibfnamefont {I.}~\bibnamefont
  {Belopolski}}, \bibinfo {author} {\bibfnamefont {S.-Y.}\ \bibnamefont {Xu}},
  \bibinfo {author} {\bibfnamefont {D.~S.}\ \bibnamefont {Sanchez}}, \bibinfo
  {author} {\bibfnamefont {G.}~\bibnamefont {Chang}}, \bibinfo {author}
  {\bibfnamefont {C.}~\bibnamefont {Guo}}, \bibinfo {author} {\bibfnamefont
  {M.}~\bibnamefont {Neupane}}, \bibinfo {author} {\bibfnamefont
  {H.}~\bibnamefont {Zheng}}, \bibinfo {author} {\bibfnamefont {C.-C.}\
  \bibnamefont {Lee}}, \bibinfo {author} {\bibfnamefont {S.-M.}\ \bibnamefont
  {Huang}}, \bibinfo {author} {\bibfnamefont {G.}~\bibnamefont {Bian}},
  \bibinfo {author} {\bibfnamefont {N.}~\bibnamefont {Alidoust}}, \bibinfo
  {author} {\bibfnamefont {T.-R.}\ \bibnamefont {Chang}}, \bibinfo {author}
  {\bibfnamefont {B.}~\bibnamefont {Wang}}, \bibinfo {author} {\bibfnamefont
  {X.}~\bibnamefont {Zhang}}, \bibinfo {author} {\bibfnamefont
  {A.}~\bibnamefont {Bansil}}, \bibinfo {author} {\bibfnamefont {H.-T.}\
  \bibnamefont {Jeng}}, \bibinfo {author} {\bibfnamefont {H.}~\bibnamefont
  {Lin}}, \bibinfo {author} {\bibfnamefont {S.}~\bibnamefont {Jia}}, \ and\
  \bibinfo {author} {\bibfnamefont {M.~Z.}\ \bibnamefont {Hasan}},\ }\href
  {\doibase 10.1103/PhysRevLett.116.066802} {\bibfield  {journal} {\bibinfo
  {journal} {Phys. Rev. Lett.}\ }\textbf {\bibinfo {volume} {116}},\ \bibinfo
  {pages} {066802} (\bibinfo {year} {2016})}\BibitemShut {NoStop}%
\bibitem [{\citenamefont {Weng}\ \emph {et~al.}()\citenamefont {Weng},
  \citenamefont {Dai},\ and\ \citenamefont {Fang}}]{Weng:2016aa}%
  \BibitemOpen
  \bibfield  {author} {\bibinfo {author} {\bibfnamefont {H.}~\bibnamefont
  {Weng}}, \bibinfo {author} {\bibfnamefont {X.}~\bibnamefont {Dai}}, \ and\
  \bibinfo {author} {\bibfnamefont {Z.}~\bibnamefont {Fang}},\ }\href
  {http://arxiv.org/abs/1603.04744} {\ }\Eprint
  {http://arxiv.org/abs/arXiv:1603.04744} {arXiv:1603.04744} \BibitemShut
  {NoStop}%
\bibitem [{\citenamefont {Huang}\ \emph {et~al.}(2015)\citenamefont {Huang},
  \citenamefont {Xu}, \citenamefont {Belopolski}, \citenamefont {Lee},
  \citenamefont {Chang}, \citenamefont {Wang}, \citenamefont {Alidoust},
  \citenamefont {Bian}, \citenamefont {Neupane}, \citenamefont {Zhang},
  \citenamefont {Jia}, \citenamefont {Bansil}, \citenamefont {Lin},\ and\
  \citenamefont {Hasan}}]{Huang:2015aa}%
  \BibitemOpen
  \bibfield  {author} {\bibinfo {author} {\bibfnamefont {S.-M.}\ \bibnamefont
  {Huang}}, \bibinfo {author} {\bibfnamefont {S.-Y.}\ \bibnamefont {Xu}},
  \bibinfo {author} {\bibfnamefont {I.}~\bibnamefont {Belopolski}}, \bibinfo
  {author} {\bibfnamefont {C.-C.}\ \bibnamefont {Lee}}, \bibinfo {author}
  {\bibfnamefont {G.}~\bibnamefont {Chang}}, \bibinfo {author} {\bibfnamefont
  {B.}~\bibnamefont {Wang}}, \bibinfo {author} {\bibfnamefont {N.}~\bibnamefont
  {Alidoust}}, \bibinfo {author} {\bibfnamefont {G.}~\bibnamefont {Bian}},
  \bibinfo {author} {\bibfnamefont {M.}~\bibnamefont {Neupane}}, \bibinfo
  {author} {\bibfnamefont {C.}~\bibnamefont {Zhang}}, \bibinfo {author}
  {\bibfnamefont {S.}~\bibnamefont {Jia}}, \bibinfo {author} {\bibfnamefont
  {A.}~\bibnamefont {Bansil}}, \bibinfo {author} {\bibfnamefont
  {H.}~\bibnamefont {Lin}}, \ and\ \bibinfo {author} {\bibfnamefont {M.~Z.}\
  \bibnamefont {Hasan}},\ }\href@noop {} {\bibfield  {journal} {\bibinfo
  {journal} {Nat. Commun.}\ }\textbf {\bibinfo {volume} {6}} (\bibinfo {year}
  {2015})}\BibitemShut {NoStop}%
\bibitem [{\citenamefont {Xu}\ \emph {et~al.}(2015{\natexlab{d}})\citenamefont
  {Xu}, \citenamefont {Du}, \citenamefont {Wang}, \citenamefont {Li},
  \citenamefont {Niu}, \citenamefont {Yao}, \citenamefont {Dudin},
  \citenamefont {Xu}, \citenamefont {Wan},\ and\ \citenamefont
  {Feng}}]{FengNbP}%
  \BibitemOpen
  \bibfield  {author} {\bibinfo {author} {\bibfnamefont {D.-F.}\ \bibnamefont
  {Xu}}, \bibinfo {author} {\bibfnamefont {Y.-P.}\ \bibnamefont {Du}}, \bibinfo
  {author} {\bibfnamefont {Z.}~\bibnamefont {Wang}}, \bibinfo {author}
  {\bibfnamefont {Y.-P.}\ \bibnamefont {Li}}, \bibinfo {author} {\bibfnamefont
  {X.-H.}\ \bibnamefont {Niu}}, \bibinfo {author} {\bibfnamefont
  {Q.}~\bibnamefont {Yao}}, \bibinfo {author} {\bibfnamefont {P.}~\bibnamefont
  {Dudin}}, \bibinfo {author} {\bibfnamefont {Z.-A.}\ \bibnamefont {Xu}},
  \bibinfo {author} {\bibfnamefont {X.-G.}\ \bibnamefont {Wan}}, \ and\
  \bibinfo {author} {\bibfnamefont {D.-L.}\ \bibnamefont {Feng}},\ }\href
  {http://stacks.iop.org/0256-307X/32/i=10/a=107101} {\bibfield  {journal}
  {\bibinfo  {journal} {Chin. Phys. Lett.}\ }\textbf {\bibinfo {volume} {32}},\
  \bibinfo {pages} {107101} (\bibinfo {year} {2015}{\natexlab{d}})},\ \Eprint
  {http://arxiv.org/abs/arXiv:1509.03847} {arXiv:1509.03847} \BibitemShut
  {NoStop}%
\bibitem [{\citenamefont {Soluyanov}\ \emph {et~al.}(2015)\citenamefont
  {Soluyanov}, \citenamefont {Gresch}, \citenamefont {Wang}, \citenamefont
  {Wu}, \citenamefont {Troyer}, \citenamefont {Dai},\ and\ \citenamefont
  {Bernevig}}]{Soluyanov:2015aa}%
  \BibitemOpen
  \bibfield  {author} {\bibinfo {author} {\bibfnamefont {A.~A.}\ \bibnamefont
  {Soluyanov}}, \bibinfo {author} {\bibfnamefont {D.}~\bibnamefont {Gresch}},
  \bibinfo {author} {\bibfnamefont {Z.}~\bibnamefont {Wang}}, \bibinfo {author}
  {\bibfnamefont {Q.}~\bibnamefont {Wu}}, \bibinfo {author} {\bibfnamefont
  {M.}~\bibnamefont {Troyer}}, \bibinfo {author} {\bibfnamefont
  {X.}~\bibnamefont {Dai}}, \ and\ \bibinfo {author} {\bibfnamefont {B.~A.}\
  \bibnamefont {Bernevig}},\ }\href {\doibase 10.1038/nature15768} {\bibfield
  {journal} {\bibinfo  {journal} {Nature}\ }\textbf {\bibinfo {volume} {527}},\
  \bibinfo {pages} {495} (\bibinfo {year} {2015})}\BibitemShut {NoStop}%
  \bibitem [{\citenamefont {Liang}\ \emph {et~al.}()\citenamefont {Liang},
  \citenamefont {Huang}, \citenamefont {Nie}, \citenamefont {Ding},
  \citenamefont {Gao}, \citenamefont {Hu}, \citenamefont {He}, \citenamefont
  {Zhang}, \citenamefont {Wang}, \citenamefont {Shen}, \citenamefont {Liu},
  \citenamefont {Ai}, \citenamefont {Yu}, \citenamefont {Sun}, \citenamefont
  {Zhao}, \citenamefont {Lv}, \citenamefont {Liu}, \citenamefont {Li},
  \citenamefont {Zhang}, \citenamefont {Hu}, \citenamefont {Xu}, \citenamefont
  {Zhao}, \citenamefont {Liu}, \citenamefont {Mao}, \citenamefont {Jia},
  \citenamefont {Zhang}, \citenamefont {Zhang}, \citenamefont {Yang},
  \citenamefont {Wang}, \citenamefont {Peng}, \citenamefont {Weng},
  \citenamefont {Dai}, \citenamefont {Fang}, \citenamefont {Xu}, \citenamefont
  {Chen},\ and\ \citenamefont {Zhou}}]{Liang:2016aa}%
  \BibitemOpen
  \bibfield  {author} {\bibinfo {author} {\bibfnamefont {A.}~\bibnamefont
  {Liang}}, \bibinfo {author} {\bibfnamefont {J.}~\bibnamefont {Huang}},
  \bibinfo {author} {\bibfnamefont {S.}~\bibnamefont {Nie}}, \bibinfo {author}
  {\bibfnamefont {Y.}~\bibnamefont {Ding}}, \bibinfo {author} {\bibfnamefont
  {Q.}~\bibnamefont {Gao}}, \bibinfo {author} {\bibfnamefont {C.}~\bibnamefont
  {Hu}}, \bibinfo {author} {\bibfnamefont {S.}~\bibnamefont {He}}, \bibinfo
  {author} {\bibfnamefont {Y.}~\bibnamefont {Zhang}}, \bibinfo {author}
  {\bibfnamefont {C.}~\bibnamefont {Wang}}, \bibinfo {author} {\bibfnamefont
  {B.}~\bibnamefont {Shen}}, \bibinfo {author} {\bibfnamefont {J.}~\bibnamefont
  {Liu}}, \bibinfo {author} {\bibfnamefont {P.}~\bibnamefont {Ai}}, \bibinfo
  {author} {\bibfnamefont {L.}~\bibnamefont {Yu}}, \bibinfo {author}
  {\bibfnamefont {X.}~\bibnamefont {Sun}}, \bibinfo {author} {\bibfnamefont
  {W.}~\bibnamefont {Zhao}}, \bibinfo {author} {\bibfnamefont {S.}~\bibnamefont
  {Lv}}, \bibinfo {author} {\bibfnamefont {D.}~\bibnamefont {Liu}}, \bibinfo
  {author} {\bibfnamefont {C.}~\bibnamefont {Li}}, \bibinfo {author}
  {\bibfnamefont {Y.}~\bibnamefont {Zhang}}, \bibinfo {author} {\bibfnamefont
  {Y.}~\bibnamefont {Hu}}, \bibinfo {author} {\bibfnamefont {Y.}~\bibnamefont
  {Xu}}, \bibinfo {author} {\bibfnamefont {L.}~\bibnamefont {Zhao}}, \bibinfo
  {author} {\bibfnamefont {G.}~\bibnamefont {Liu}}, \bibinfo {author}
  {\bibfnamefont {Z.}~\bibnamefont {Mao}}, \bibinfo {author} {\bibfnamefont
  {X.}~\bibnamefont {Jia}}, \bibinfo {author} {\bibfnamefont {F.}~\bibnamefont
  {Zhang}}, \bibinfo {author} {\bibfnamefont {S.}~\bibnamefont {Zhang}},
  \bibinfo {author} {\bibfnamefont {F.}~\bibnamefont {Yang}}, \bibinfo {author}
  {\bibfnamefont {Z.}~\bibnamefont {Wang}}, \bibinfo {author} {\bibfnamefont
  {Q.}~\bibnamefont {Peng}}, \bibinfo {author} {\bibfnamefont {H.}~\bibnamefont
  {Weng}}, \bibinfo {author} {\bibfnamefont {X.}~\bibnamefont {Dai}}, \bibinfo
  {author} {\bibfnamefont {Z.}~\bibnamefont {Fang}}, \bibinfo {author}
  {\bibfnamefont {Z.}~\bibnamefont {Xu}}, \bibinfo {author} {\bibfnamefont
  {C.}~\bibnamefont {Chen}}, \ and\ \bibinfo {author} {\bibfnamefont {X.~J.}\
  \bibnamefont {Zhou}},\ }\href {http://arxiv.org/abs/1604.01706} {\ }\Eprint
  {http://arxiv.org/abs/arXiv:1604.01706} {arXiv:1604.01706} \BibitemShut
  {NoStop}%
\bibitem [{\citenamefont {{Borisenko}}\ \emph {et~al.}(2015)\citenamefont
  {{Borisenko}}, \citenamefont {{Evtushinsky}}, \citenamefont {{Gibson}},
  \citenamefont {{Yaresko}}, \citenamefont {{Kim}}, \citenamefont {{Ali}},
  \citenamefont {{Buechner}}, \citenamefont {{Hoesch}},\ and\ \citenamefont
  {{Cava}}}]{type2a}%
  \BibitemOpen
  \bibfield  {author} {\bibinfo {author} {\bibfnamefont {S.}~\bibnamefont
  {{Borisenko}}}, \bibinfo {author} {\bibfnamefont {D.}~\bibnamefont
  {{Evtushinsky}}}, \bibinfo {author} {\bibfnamefont {Q.}~\bibnamefont
  {{Gibson}}}, \bibinfo {author} {\bibfnamefont {A.}~\bibnamefont {{Yaresko}}},
  \bibinfo {author} {\bibfnamefont {T.}~\bibnamefont {{Kim}}}, \bibinfo
  {author} {\bibfnamefont {M.~N.}\ \bibnamefont {{Ali}}}, \bibinfo {author}
  {\bibfnamefont {B.}~\bibnamefont {{Buechner}}}, \bibinfo {author}
  {\bibfnamefont {M.}~\bibnamefont {{Hoesch}}}, \ and\ \bibinfo {author}
  {\bibfnamefont {R.~J.}\ \bibnamefont {{Cava}}},\ }\href@noop {} {\bibfield
  {journal} {\bibinfo  {journal} {ArXiv e-prints}\ } (\bibinfo {year}
  {2015})},\ \Eprint {http://arxiv.org/abs/1507.04847} {arXiv:1507.04847
  [cond-mat.mes-hall]} \BibitemShut {NoStop}%
\bibitem [{\citenamefont {{Wang}}\ \emph
  {et~al.}(2016{\natexlab{a}})\citenamefont {{Wang}}, \citenamefont {{Zhang}},
  \citenamefont {{Huang}}, \citenamefont {{Nie}}, \citenamefont {{Liu}},
  \citenamefont {{Liang}}, \citenamefont {{Zhang}}, \citenamefont {{Shen}},
  \citenamefont {{Liu}}, \citenamefont {{Hu}}, \citenamefont {{Ding}},
  \citenamefont {{Liu}}, \citenamefont {{Hu}}, \citenamefont {{He}},
  \citenamefont {{Zhao}}, \citenamefont {{Yu}}, \citenamefont {{Hu}},
  \citenamefont {{Wei}}, \citenamefont {{Mao}}, \citenamefont {{Shi}},
  \citenamefont {{Jia}}, \citenamefont {{Zhang}}, \citenamefont {{Zhang}},
  \citenamefont {{Yang}}, \citenamefont {{Wang}}, \citenamefont {{Peng}},
  \citenamefont {{Weng}}, \citenamefont {{Dai}}, \citenamefont {{Fang}},
  \citenamefont {{Xu}}, \citenamefont {{Chen}},\ and\ \citenamefont
  {{Zhou}}}]{type2b}%
  \BibitemOpen
  \bibfield  {author} {\bibinfo {author} {\bibfnamefont {C.}~\bibnamefont
  {{Wang}}}, \bibinfo {author} {\bibfnamefont {Y.}~\bibnamefont {{Zhang}}},
  \bibinfo {author} {\bibfnamefont {J.}~\bibnamefont {{Huang}}}, \bibinfo
  {author} {\bibfnamefont {S.}~\bibnamefont {{Nie}}}, \bibinfo {author}
  {\bibfnamefont {G.}~\bibnamefont {{Liu}}}, \bibinfo {author} {\bibfnamefont
  {A.}~\bibnamefont {{Liang}}}, \bibinfo {author} {\bibfnamefont
  {Y.}~\bibnamefont {{Zhang}}}, \bibinfo {author} {\bibfnamefont
  {B.}~\bibnamefont {{Shen}}}, \bibinfo {author} {\bibfnamefont
  {J.}~\bibnamefont {{Liu}}}, \bibinfo {author} {\bibfnamefont
  {C.}~\bibnamefont {{Hu}}}, \bibinfo {author} {\bibfnamefont {Y.}~\bibnamefont
  {{Ding}}}, \bibinfo {author} {\bibfnamefont {D.}~\bibnamefont {{Liu}}},
  \bibinfo {author} {\bibfnamefont {Y.}~\bibnamefont {{Hu}}}, \bibinfo {author}
  {\bibfnamefont {S.}~\bibnamefont {{He}}}, \bibinfo {author} {\bibfnamefont
  {L.}~\bibnamefont {{Zhao}}}, \bibinfo {author} {\bibfnamefont
  {L.}~\bibnamefont {{Yu}}}, \bibinfo {author} {\bibfnamefont {J.}~\bibnamefont
  {{Hu}}}, \bibinfo {author} {\bibfnamefont {J.}~\bibnamefont {{Wei}}},
  \bibinfo {author} {\bibfnamefont {Z.}~\bibnamefont {{Mao}}}, \bibinfo
  {author} {\bibfnamefont {Y.}~\bibnamefont {{Shi}}}, \bibinfo {author}
  {\bibfnamefont {X.}~\bibnamefont {{Jia}}}, \bibinfo {author} {\bibfnamefont
  {F.}~\bibnamefont {{Zhang}}}, \bibinfo {author} {\bibfnamefont
  {S.}~\bibnamefont {{Zhang}}}, \bibinfo {author} {\bibfnamefont
  {F.}~\bibnamefont {{Yang}}}, \bibinfo {author} {\bibfnamefont
  {Z.}~\bibnamefont {{Wang}}}, \bibinfo {author} {\bibfnamefont
  {Q.}~\bibnamefont {{Peng}}}, \bibinfo {author} {\bibfnamefont
  {H.}~\bibnamefont {{Weng}}}, \bibinfo {author} {\bibfnamefont
  {X.}~\bibnamefont {{Dai}}}, \bibinfo {author} {\bibfnamefont
  {Z.}~\bibnamefont {{Fang}}}, \bibinfo {author} {\bibfnamefont
  {Z.}~\bibnamefont {{Xu}}}, \bibinfo {author} {\bibfnamefont {C.}~\bibnamefont
  {{Chen}}}, \ and\ \bibinfo {author} {\bibfnamefont {X.~J.}\ \bibnamefont
  {{Zhou}}},\ }\href@noop {} {\bibfield  {journal} {\bibinfo  {journal} {ArXiv
  e-prints}\ } (\bibinfo {year} {2016}{\natexlab{a}})},\ \Eprint
  {http://arxiv.org/abs/1604.04218} {arXiv:1604.04218 [cond-mat.mes-hall]}
  \BibitemShut {NoStop}%
\bibitem [{\citenamefont {{Jiang}}\ \emph {et~al.}(2016)\citenamefont
  {{Jiang}}, \citenamefont {{Liu}}, \citenamefont {{Sun}}, \citenamefont
  {{Yang}}, \citenamefont {{Rajamathi}}, \citenamefont {{Qi}}, \citenamefont
  {{Yang}}, \citenamefont {{Chen}}, \citenamefont {{Peng}}, \citenamefont
  {{Hwang}}, \citenamefont {{Sun}}, \citenamefont {{Mo}}, \citenamefont
  {{Vobornik}}, \citenamefont {{Fujii}}, \citenamefont {{Parkin}},
  \citenamefont {{Felser}}, \citenamefont {{Yan}},\ and\ \citenamefont
  {{Chen}}}]{type2c}%
  \BibitemOpen
  \bibfield  {author} {\bibinfo {author} {\bibfnamefont {J.}~\bibnamefont
  {{Jiang}}}, \bibinfo {author} {\bibfnamefont {Z.~K.}\ \bibnamefont {{Liu}}},
  \bibinfo {author} {\bibfnamefont {Y.}~\bibnamefont {{Sun}}}, \bibinfo
  {author} {\bibfnamefont {H.~F.}\ \bibnamefont {{Yang}}}, \bibinfo {author}
  {\bibfnamefont {R.}~\bibnamefont {{Rajamathi}}}, \bibinfo {author}
  {\bibfnamefont {Y.~P.}\ \bibnamefont {{Qi}}}, \bibinfo {author}
  {\bibfnamefont {L.~X.}\ \bibnamefont {{Yang}}}, \bibinfo {author}
  {\bibfnamefont {C.}~\bibnamefont {{Chen}}}, \bibinfo {author} {\bibfnamefont
  {H.}~\bibnamefont {{Peng}}}, \bibinfo {author} {\bibfnamefont {C.-C.}\
  \bibnamefont {{Hwang}}}, \bibinfo {author} {\bibfnamefont {S.~Z.}\
  \bibnamefont {{Sun}}}, \bibinfo {author} {\bibfnamefont {S.-K.}\ \bibnamefont
  {{Mo}}}, \bibinfo {author} {\bibfnamefont {I.}~\bibnamefont {{Vobornik}}},
  \bibinfo {author} {\bibfnamefont {J.}~\bibnamefont {{Fujii}}}, \bibinfo
  {author} {\bibfnamefont {S.~S.~P.}\ \bibnamefont {{Parkin}}}, \bibinfo
  {author} {\bibfnamefont {C.}~\bibnamefont {{Felser}}}, \bibinfo {author}
  {\bibfnamefont {B.~H.}\ \bibnamefont {{Yan}}}, \ and\ \bibinfo {author}
  {\bibfnamefont {Y.~L.}\ \bibnamefont {{Chen}}},\ }\href@noop {} {\bibfield
  {journal} {\bibinfo  {journal} {ArXiv e-prints}\ } (\bibinfo {year}
  {2016})},\ \Eprint {http://arxiv.org/abs/1604.00139} {arXiv:1604.00139
  [cond-mat.mtrl-sci]} \BibitemShut {NoStop}%
\bibitem [{\citenamefont {{Deng}}\ \emph {et~al.}(2016)\citenamefont {{Deng}},
  \citenamefont {{Wan}}, \citenamefont {{Deng}}, \citenamefont {{Zhang}},
  \citenamefont {{Ding}}, \citenamefont {{Wang}}, \citenamefont {{Yan}},
  \citenamefont {{Huang}}, \citenamefont {{Zhang}}, \citenamefont {{Xu}},
  \citenamefont {{Denlinger}}, \citenamefont {{Fedorov}}, \citenamefont
  {{Yang}}, \citenamefont {{Duan}}, \citenamefont {{Yao}}, \citenamefont
  {{Wu}}, \citenamefont {{Shoushan Fan}}, \citenamefont {{Zhang}},
  \citenamefont {{Chen}},\ and\ \citenamefont {{Zhou}}}]{type2d}%
  \BibitemOpen
  \bibfield  {author} {\bibinfo {author} {\bibfnamefont {K.}~\bibnamefont
  {{Deng}}}, \bibinfo {author} {\bibfnamefont {G.}~\bibnamefont {{Wan}}},
  \bibinfo {author} {\bibfnamefont {P.}~\bibnamefont {{Deng}}}, \bibinfo
  {author} {\bibfnamefont {K.}~\bibnamefont {{Zhang}}}, \bibinfo {author}
  {\bibfnamefont {S.}~\bibnamefont {{Ding}}}, \bibinfo {author} {\bibfnamefont
  {E.}~\bibnamefont {{Wang}}}, \bibinfo {author} {\bibfnamefont
  {M.}~\bibnamefont {{Yan}}}, \bibinfo {author} {\bibfnamefont
  {H.}~\bibnamefont {{Huang}}}, \bibinfo {author} {\bibfnamefont
  {H.}~\bibnamefont {{Zhang}}}, \bibinfo {author} {\bibfnamefont
  {Z.}~\bibnamefont {{Xu}}}, \bibinfo {author} {\bibfnamefont {J.}~\bibnamefont
  {{Denlinger}}}, \bibinfo {author} {\bibfnamefont {A.}~\bibnamefont
  {{Fedorov}}}, \bibinfo {author} {\bibfnamefont {H.}~\bibnamefont {{Yang}}},
  \bibinfo {author} {\bibfnamefont {W.}~\bibnamefont {{Duan}}}, \bibinfo
  {author} {\bibfnamefont {H.}~\bibnamefont {{Yao}}}, \bibinfo {author}
  {\bibfnamefont {Y.}~\bibnamefont {{Wu}}}, \bibinfo {author} {\bibfnamefont
  {y.}~\bibnamefont {{Shoushan Fan}}}, \bibinfo {author} {\bibfnamefont
  {H.}~\bibnamefont {{Zhang}}}, \bibinfo {author} {\bibfnamefont
  {X.}~\bibnamefont {{Chen}}}, \ and\ \bibinfo {author} {\bibfnamefont
  {S.}~\bibnamefont {{Zhou}}},\ }\href@noop {} {\bibfield  {journal} {\bibinfo
  {journal} {ArXiv e-prints}\ } (\bibinfo {year} {2016})},\ \Eprint
  {http://arxiv.org/abs/1603.08508} {arXiv:1603.08508 [cond-mat.mes-hall]}
  \BibitemShut {NoStop}%
  \bibitem{bhyanPRB2015}Y. Sun, S.-C. Wu, and B. Yan
Phys. Rev. B \textbf{92}, 115428 (2015).
\bibitem [{\citenamefont {{Bruno}}\ \emph {et~al.}(2016)\citenamefont
  {{Bruno}}, \citenamefont {{Tamai}}, \citenamefont {{Wu}}, \citenamefont
  {{Cucchi}}, \citenamefont {{Barreteau}}, \citenamefont {{de la Torre}},
  \citenamefont {{McKeown Walker}}, \citenamefont {{Ricc{\`o}}}, \citenamefont
  {{Wang}}, \citenamefont {{Kim}}, \citenamefont {{Hoesch}}, \citenamefont
  {{Shi}}, \citenamefont {{Plumb}}, \citenamefont {{Giannini}}, \citenamefont
  {{Soluyanov}},\ and\ \citenamefont {{Baumberger}}}]{type2e}%
  \BibitemOpen
  \bibfield  {author} {\bibinfo {author} {\bibfnamefont {F.~Y.}\ \bibnamefont
  {{Bruno}}}, \bibinfo {author} {\bibfnamefont {A.}~\bibnamefont {{Tamai}}},
  \bibinfo {author} {\bibfnamefont {Q.~S.}\ \bibnamefont {{Wu}}}, \bibinfo
  {author} {\bibfnamefont {I.}~\bibnamefont {{Cucchi}}}, \bibinfo {author}
  {\bibfnamefont {C.}~\bibnamefont {{Barreteau}}}, \bibinfo {author}
  {\bibfnamefont {A.}~\bibnamefont {{de la Torre}}}, \bibinfo {author}
  {\bibfnamefont {S.}~\bibnamefont {{McKeown Walker}}}, \bibinfo {author}
  {\bibfnamefont {S.}~\bibnamefont {{Ricc{\`o}}}}, \bibinfo {author}
  {\bibfnamefont {Z.}~\bibnamefont {{Wang}}}, \bibinfo {author} {\bibfnamefont
  {T.~K.}\ \bibnamefont {{Kim}}}, \bibinfo {author} {\bibfnamefont
  {M.}~\bibnamefont {{Hoesch}}}, \bibinfo {author} {\bibfnamefont
  {M.}~\bibnamefont {{Shi}}}, \bibinfo {author} {\bibfnamefont {N.~C.}\
  \bibnamefont {{Plumb}}}, \bibinfo {author} {\bibfnamefont {E.}~\bibnamefont
  {{Giannini}}}, \bibinfo {author} {\bibfnamefont {A.~A.}\ \bibnamefont
  {{Soluyanov}}}, \ and\ \bibinfo {author} {\bibfnamefont {F.}~\bibnamefont
  {{Baumberger}}},\ }\href@noop {} {\bibfield  {journal} {\bibinfo  {journal}
  {ArXiv e-prints}\ } (\bibinfo {year} {2016})},\ \Eprint
  {http://arxiv.org/abs/1604.02411} {arXiv:1604.02411 [cond-mat.mtrl-sci]}
  \BibitemShut {NoStop}%
\bibitem [{\citenamefont {Wang}\ \emph {et~al.}(2012)\citenamefont {Wang},
  \citenamefont {Sun}, \citenamefont {Chen}, \citenamefont {Franchini},
  \citenamefont {Xu}, \citenamefont {Weng}, \citenamefont {Dai},\ and\
  \citenamefont {Fang}}]{Wang_Dirac2012}%
  \BibitemOpen
  \bibfield  {author} {\bibinfo {author} {\bibfnamefont {Z.}~\bibnamefont
  {Wang}}, \bibinfo {author} {\bibfnamefont {Y.}~\bibnamefont {Sun}}, \bibinfo
  {author} {\bibfnamefont {X.-Q.}\ \bibnamefont {Chen}}, \bibinfo {author}
  {\bibfnamefont {C.}~\bibnamefont {Franchini}}, \bibinfo {author}
  {\bibfnamefont {G.}~\bibnamefont {Xu}}, \bibinfo {author} {\bibfnamefont
  {H.}~\bibnamefont {Weng}}, \bibinfo {author} {\bibfnamefont {X.}~\bibnamefont
  {Dai}}, \ and\ \bibinfo {author} {\bibfnamefont {Z.}~\bibnamefont {Fang}},\
  }\href {\doibase 10.1103/PhysRevB.85.195320} {\bibfield  {journal} {\bibinfo
  {journal} {Phys. Rev. B}\ }\textbf {\bibinfo {volume} {85}},\ \bibinfo
  {pages} {195320} (\bibinfo {year} {2012})}\BibitemShut {NoStop}%
\bibitem [{\citenamefont {Wang}\ \emph {et~al.}(2013)\citenamefont {Wang},
  \citenamefont {Weng}, \citenamefont {Wu}, \citenamefont {Dai},\ and\
  \citenamefont {Fang}}]{Wang_Dirac2013}%
  \BibitemOpen
  \bibfield  {author} {\bibinfo {author} {\bibfnamefont {Z.}~\bibnamefont
  {Wang}}, \bibinfo {author} {\bibfnamefont {H.}~\bibnamefont {Weng}}, \bibinfo
  {author} {\bibfnamefont {Q.}~\bibnamefont {Wu}}, \bibinfo {author}
  {\bibfnamefont {X.}~\bibnamefont {Dai}}, \ and\ \bibinfo {author}
  {\bibfnamefont {Z.}~\bibnamefont {Fang}},\ }\href {\doibase
  10.1103/PhysRevB.88.125427} {\bibfield  {journal} {\bibinfo  {journal} {Phys.
  Rev. B}\ }\textbf {\bibinfo {volume} {88}},\ \bibinfo {pages} {125427}
  (\bibinfo {year} {2013})}\BibitemShut {NoStop}%
\bibitem [{\citenamefont {Liu}\ \emph {et~al.}(2014)\citenamefont {Liu},
  \citenamefont {Zhou}, \citenamefont {Zhang}, \citenamefont {Wang},
  \citenamefont {Weng}, \citenamefont {Prabhakaran}, \citenamefont {Mo},
  \citenamefont {Shen}, \citenamefont {Fang}, \citenamefont {Dai},
  \citenamefont {Hussain},\ and\ \citenamefont {Chen}}]{Liu864}%
  \BibitemOpen
  \bibfield  {author} {\bibinfo {author} {\bibfnamefont {Z.~K.}\ \bibnamefont
  {Liu}}, \bibinfo {author} {\bibfnamefont {B.}~\bibnamefont {Zhou}}, \bibinfo
  {author} {\bibfnamefont {Y.}~\bibnamefont {Zhang}}, \bibinfo {author}
  {\bibfnamefont {Z.~J.}\ \bibnamefont {Wang}}, \bibinfo {author}
  {\bibfnamefont {H.~M.}\ \bibnamefont {Weng}}, \bibinfo {author}
  {\bibfnamefont {D.}~\bibnamefont {Prabhakaran}}, \bibinfo {author}
  {\bibfnamefont {S.-K.}\ \bibnamefont {Mo}}, \bibinfo {author} {\bibfnamefont
  {Z.~X.}\ \bibnamefont {Shen}}, \bibinfo {author} {\bibfnamefont
  {Z.}~\bibnamefont {Fang}}, \bibinfo {author} {\bibfnamefont {X.}~\bibnamefont
  {Dai}}, \bibinfo {author} {\bibfnamefont {Z.}~\bibnamefont {Hussain}}, \ and\
  \bibinfo {author} {\bibfnamefont {Y.~L.}\ \bibnamefont {Chen}},\ }\href
  {\doibase 10.1126/science.1245085} {\bibfield  {journal} {\bibinfo  {journal}
  {Science}\ }\textbf {\bibinfo {volume} {343}},\ \bibinfo {pages} {864}
  (\bibinfo {year} {2014})}\BibitemShut {NoStop}%
\bibitem [{\citenamefont {Borisenko}\ \emph {et~al.}(2014)\citenamefont
  {Borisenko}, \citenamefont {Gibson}, \citenamefont {Evtushinsky},
  \citenamefont {Zabolotnyy}, \citenamefont {B\"uchner},\ and\ \citenamefont
  {Cava}}]{cava2014}%
  \BibitemOpen
  \bibfield  {author} {\bibinfo {author} {\bibfnamefont {S.}~\bibnamefont
  {Borisenko}}, \bibinfo {author} {\bibfnamefont {Q.}~\bibnamefont {Gibson}},
  \bibinfo {author} {\bibfnamefont {D.}~\bibnamefont {Evtushinsky}}, \bibinfo
  {author} {\bibfnamefont {V.}~\bibnamefont {Zabolotnyy}}, \bibinfo {author}
  {\bibfnamefont {B.}~\bibnamefont {B\"uchner}}, \ and\ \bibinfo {author}
  {\bibfnamefont {R.~J.}\ \bibnamefont {Cava}},\ }\href {\doibase
  10.1103/PhysRevLett.113.027603} {\bibfield  {journal} {\bibinfo  {journal}
  {Phys. Rev. Lett.}\ }\textbf {\bibinfo {volume} {113}},\ \bibinfo {pages}
  {027603} (\bibinfo {year} {2014})}\BibitemShut {NoStop}%
\bibitem [{\citenamefont {Zhang}\ \emph {et~al.}(2016)\citenamefont {Zhang},
  \citenamefont {Hutasoit}, \citenamefont {Sun}, \citenamefont {Yan},
  \citenamefont {Xu},\ and\ \citenamefont {Liu}}]{nem_dirac}%
  \BibitemOpen
  \bibfield  {author} {\bibinfo {author} {\bibfnamefont {R.-X.}\ \bibnamefont
  {Zhang}}, \bibinfo {author} {\bibfnamefont {J.~A.}\ \bibnamefont {Hutasoit}},
  \bibinfo {author} {\bibfnamefont {Y.}~\bibnamefont {Sun}}, \bibinfo {author}
  {\bibfnamefont {B.}~\bibnamefont {Yan}}, \bibinfo {author} {\bibfnamefont
  {C.}~\bibnamefont {Xu}}, \ and\ \bibinfo {author} {\bibfnamefont {C.-X.}\
  \bibnamefont {Liu}},\ }\href {\doibase 10.1103/PhysRevB.93.041108} {\bibfield
   {journal} {\bibinfo  {journal} {Phys. Rev. B}\ }\textbf {\bibinfo {volume}
  {93}},\ \bibinfo {pages} {041108} (\bibinfo {year} {2016})}\BibitemShut
  {NoStop}%
\bibitem [{\citenamefont {{Moll}}\ \emph {et~al.}(2015)\citenamefont {{Moll}},
  \citenamefont {{Nair}}, \citenamefont {{Helm}}, \citenamefont {{Potter}},
  \citenamefont {{Kimchi}}, \citenamefont {{Vishwanath}},\ and\ \citenamefont
  {{Analytis}}}]{jim2015}%
  \BibitemOpen
  \bibfield  {author} {\bibinfo {author} {\bibfnamefont {P.~J.~W.}\
  \bibnamefont {{Moll}}}, \bibinfo {author} {\bibfnamefont {N.~L.}\
  \bibnamefont {{Nair}}}, \bibinfo {author} {\bibfnamefont {T.}~\bibnamefont
  {{Helm}}}, \bibinfo {author} {\bibfnamefont {A.~C.}\ \bibnamefont
  {{Potter}}}, \bibinfo {author} {\bibfnamefont {I.}~\bibnamefont {{Kimchi}}},
  \bibinfo {author} {\bibfnamefont {A.}~\bibnamefont {{Vishwanath}}}, \ and\
  \bibinfo {author} {\bibfnamefont {J.~G.}\ \bibnamefont {{Analytis}}},\
  }\href@noop {} {\bibfield  {journal} {\bibinfo  {journal} {ArXiv e-prints}\ }
  (\bibinfo {year} {2015})},\ \Eprint {http://arxiv.org/abs/1505.02817}
  {arXiv:1505.02817 [cond-mat.mes-hall]} \BibitemShut {NoStop}%
\bibitem [{\citenamefont {Nielsen}\ and\ \citenamefont
  {Ninomiya}(1981{\natexlab{a}})}]{NIELSEN198120}%
  \BibitemOpen
  \bibfield  {author} {\bibinfo {author} {\bibfnamefont {H.}~\bibnamefont
  {Nielsen}}\ and\ \bibinfo {author} {\bibfnamefont {M.}~\bibnamefont
  {Ninomiya}},\ }\href {\doibase
  http://dx.doi.org/10.1016/0550-3213(81)90361-8} {\bibfield  {journal}
  {\bibinfo  {journal} {Nucl. Phys. B}\ }\textbf {\bibinfo {volume} {185}},\
  \bibinfo {pages} {20 } (\bibinfo {year} {1981}{\natexlab{a}})}\BibitemShut
  {NoStop}%
\bibitem [{\citenamefont {Nielsen}\ and\ \citenamefont
  {Ninomiya}(1981{\natexlab{b}})}]{NIELSEN1981173}%
  \BibitemOpen
  \bibfield  {author} {\bibinfo {author} {\bibfnamefont {H.}~\bibnamefont
  {Nielsen}}\ and\ \bibinfo {author} {\bibfnamefont {M.}~\bibnamefont
  {Ninomiya}},\ }\href {\doibase
  http://dx.doi.org/10.1016/0550-3213(81)90524-1} {\bibfield  {journal}
  {\bibinfo  {journal} {Nucl. Phys. B}\ }\textbf {\bibinfo {volume} {193}},\
  \bibinfo {pages} {173 } (\bibinfo {year} {1981}{\natexlab{b}})}\BibitemShut
  {NoStop}%
\bibitem [{\citenamefont {Aji}(2012)}]{aji2012}%
  \BibitemOpen
  \bibfield  {author} {\bibinfo {author} {\bibfnamefont {V.}~\bibnamefont
  {Aji}},\ }\href {\doibase 10.1103/PhysRevB.85.241101} {\bibfield  {journal}
  {\bibinfo  {journal} {Phys. Rev. B}\ }\textbf {\bibinfo {volume} {85}},\
  \bibinfo {pages} {241101} (\bibinfo {year} {2012})}\BibitemShut {NoStop}%
\bibitem [{\citenamefont {Liu}\ \emph {et~al.}(2013)\citenamefont {Liu},
  \citenamefont {Ye},\ and\ \citenamefont {Qi}}]{12a}%
  \BibitemOpen
  \bibfield  {author} {\bibinfo {author} {\bibfnamefont {C.-X.}\ \bibnamefont
  {Liu}}, \bibinfo {author} {\bibfnamefont {P.}~\bibnamefont {Ye}}, \ and\
  \bibinfo {author} {\bibfnamefont {X.-L.}\ \bibnamefont {Qi}},\ }\href
  {\doibase 10.1103/PhysRevB.87.235306} {\bibfield  {journal} {\bibinfo
  {journal} {Phys. Rev. B}\ }\textbf {\bibinfo {volume} {87}},\ \bibinfo
  {pages} {235306} (\bibinfo {year} {2013})},\ \Eprint
  {http://arxiv.org/abs/arXiv:1204.6551} {arXiv:1204.6551} \BibitemShut
  {NoStop}%
\bibitem [{\citenamefont {Son}\ and\ \citenamefont
  {Yamamoto}(2012)}]{Son_2012}%
  \BibitemOpen
  \bibfield  {author} {\bibinfo {author} {\bibfnamefont {D.~T.}\ \bibnamefont
  {Son}}\ and\ \bibinfo {author} {\bibfnamefont {N.}~\bibnamefont {Yamamoto}},\
  }\href {\doibase 10.1103/PhysRevLett.109.181602} {\bibfield  {journal}
  {\bibinfo  {journal} {Phys. Rev. Lett.}\ }\textbf {\bibinfo {volume} {109}},\
  \bibinfo {pages} {181602} (\bibinfo {year} {2012})}\BibitemShut {NoStop}%
\bibitem [{\citenamefont {Grushin}(2012)}]{Grushin_2012}%
  \BibitemOpen
  \bibfield  {author} {\bibinfo {author} {\bibfnamefont {A.~G.}\ \bibnamefont
  {Grushin}},\ }\href {\doibase 10.1103/PhysRevD.86.045001} {\bibfield
  {journal} {\bibinfo  {journal} {Phys. Rev. D}\ }\textbf {\bibinfo {volume}
  {86}},\ \bibinfo {pages} {045001} (\bibinfo {year} {2012})}\BibitemShut
  {NoStop}%
\bibitem [{\citenamefont {Zyuzin}\ and\ \citenamefont
  {Burkov}(2012)}]{burkov_theta}%
  \BibitemOpen
  \bibfield  {author} {\bibinfo {author} {\bibfnamefont {A.~A.}\ \bibnamefont
  {Zyuzin}}\ and\ \bibinfo {author} {\bibfnamefont {A.~A.}\ \bibnamefont
  {Burkov}},\ }\href {\doibase 10.1103/PhysRevB.86.115133} {\bibfield
  {journal} {\bibinfo  {journal} {Phys. Rev. B}\ }\textbf {\bibinfo {volume}
  {86}},\ \bibinfo {pages} {115133} (\bibinfo {year} {2012})}\BibitemShut
  {NoStop}%
\bibitem [{\citenamefont {Hosur}\ and\ \citenamefont {Qi}(2013)}]{hosur2013}%
  \BibitemOpen
  \bibfield  {author} {\bibinfo {author} {\bibfnamefont {P.}~\bibnamefont
  {Hosur}}\ and\ \bibinfo {author} {\bibfnamefont {X.}~\bibnamefont {Qi}},\
  }\href@noop {} {\bibfield  {journal} {\bibinfo  {journal} {Comptes Rendus
  Physique}\ }\textbf {\bibinfo {volume} {14}},\ \bibinfo {pages} {857}
  (\bibinfo {year} {2013})},\ \Eprint {http://arxiv.org/abs/arXiv:1309.4464}
  {arXiv:1309.4464} \BibitemShut {NoStop}%
\bibitem [{\citenamefont {Goswami}\ and\ \citenamefont
  {Tewari}(2013)}]{goswami2013}%
  \BibitemOpen
  \bibfield  {author} {\bibinfo {author} {\bibfnamefont {P.}~\bibnamefont
  {Goswami}}\ and\ \bibinfo {author} {\bibfnamefont {S.}~\bibnamefont
  {Tewari}},\ }\href {\doibase 10.1103/PhysRevB.88.245107} {\bibfield
  {journal} {\bibinfo  {journal} {Phys. Rev. B}\ }\textbf {\bibinfo {volume}
  {88}},\ \bibinfo {pages} {245107} (\bibinfo {year} {2013})}\BibitemShut
  {NoStop}%
\bibitem [{\citenamefont {Bednik}\ \emph {et~al.}(2015)\citenamefont {Bednik},
  \citenamefont {Zyuzin},\ and\ \citenamefont {Burkov}}]{Bednik2015}%
  \BibitemOpen
  \bibfield  {author} {\bibinfo {author} {\bibfnamefont {G.}~\bibnamefont
  {Bednik}}, \bibinfo {author} {\bibfnamefont {A.~A.}\ \bibnamefont {Zyuzin}},
  \ and\ \bibinfo {author} {\bibfnamefont {A.~A.}\ \bibnamefont {Burkov}},\
  }\href {\doibase 10.1103/PhysRevB.92.035153} {\bibfield  {journal} {\bibinfo
  {journal} {Phys. Rev. B}\ }\textbf {\bibinfo {volume} {92}},\ \bibinfo
  {pages} {035153} (\bibinfo {year} {2015})}\BibitemShut {NoStop}%
\bibitem [{\citenamefont {Lu}\ \emph {et~al.}(2015)\citenamefont {Lu},
  \citenamefont {Yada}, \citenamefont {Sato},\ and\ \citenamefont
  {Tanaka}}]{LuTanaka2015}%
  \BibitemOpen
  \bibfield  {author} {\bibinfo {author} {\bibfnamefont {B.}~\bibnamefont
  {Lu}}, \bibinfo {author} {\bibfnamefont {K.}~\bibnamefont {Yada}}, \bibinfo
  {author} {\bibfnamefont {M.}~\bibnamefont {Sato}}, \ and\ \bibinfo {author}
  {\bibfnamefont {Y.}~\bibnamefont {Tanaka}},\ }\href {\doibase
  10.1103/PhysRevLett.114.096804} {\bibfield  {journal} {\bibinfo  {journal}
  {Phys. Rev. Lett.}\ }\textbf {\bibinfo {volume} {114}},\ \bibinfo {pages}
  {096804} (\bibinfo {year} {2015})}\BibitemShut {NoStop}%
\bibitem [{\citenamefont {Li}\ and\ \citenamefont
  {Haldane}()}]{Lihaldane:2015aa}%
  \BibitemOpen
  \bibfield  {author} {\bibinfo {author} {\bibfnamefont {Y.}~\bibnamefont
  {Li}}\ and\ \bibinfo {author} {\bibfnamefont {F.~D.~M.}\ \bibnamefont
  {Haldane}},\ }\href@noop {} {\ }\Eprint
  {http://arxiv.org/abs/arXiv:1510.01730} {arXiv:1510.01730} \BibitemShut
  {NoStop}%
\bibitem [{\citenamefont {Wang}\ \emph {et~al.}()\citenamefont {Wang},
  \citenamefont {Hao}, \citenamefont {Wang},\ and\ \citenamefont
  {Ting}}]{Wang:2016aa}%
  \BibitemOpen
  \bibfield  {author} {\bibinfo {author} {\bibfnamefont {R.}~\bibnamefont
  {Wang}}, \bibinfo {author} {\bibfnamefont {L.}~\bibnamefont {Hao}}, \bibinfo
  {author} {\bibfnamefont {B.}~\bibnamefont {Wang}}, \ and\ \bibinfo {author}
  {\bibfnamefont {C.~S.}\ \bibnamefont {Ting}},\ }\href@noop {} {\ }\Eprint
  {http://arxiv.org/abs/arXiv:1602.05138} {arXiv:1602.05138} \BibitemShut
  {NoStop}%
\bibitem [{\citenamefont {Cho}\ \emph {et~al.}(2012)\citenamefont {Cho},
  \citenamefont {Bardarson}, \citenamefont {Lu},\ and\ \citenamefont
  {Moore}}]{Cho_weyl_2012}%
  \BibitemOpen
  \bibfield  {author} {\bibinfo {author} {\bibfnamefont {G.~Y.}\ \bibnamefont
  {Cho}}, \bibinfo {author} {\bibfnamefont {J.~H.}\ \bibnamefont {Bardarson}},
  \bibinfo {author} {\bibfnamefont {Y.-M.}\ \bibnamefont {Lu}}, \ and\ \bibinfo
  {author} {\bibfnamefont {J.~E.}\ \bibnamefont {Moore}},\ }\href {\doibase
  10.1103/PhysRevB.86.214514} {\bibfield  {journal} {\bibinfo  {journal} {Phys.
  Rev. B}\ }\textbf {\bibinfo {volume} {86}},\ \bibinfo {pages} {214514}
  (\bibinfo {year} {2012})}\BibitemShut {NoStop}%
\bibitem [{\citenamefont {Jian}\ \emph {et~al.}(2015)\citenamefont {Jian},
  \citenamefont {Jiang},\ and\ \citenamefont {Yao}}]{susy_weyl}%
  \BibitemOpen
  \bibfield  {author} {\bibinfo {author} {\bibfnamefont {S.-K.}\ \bibnamefont
  {Jian}}, \bibinfo {author} {\bibfnamefont {Y.-F.}\ \bibnamefont {Jiang}}, \
  and\ \bibinfo {author} {\bibfnamefont {H.}~\bibnamefont {Yao}},\ }\href
  {\doibase 10.1103/PhysRevLett.114.237001} {\bibfield  {journal} {\bibinfo
  {journal} {Phys. Rev. Lett.}\ }\textbf {\bibinfo {volume} {114}},\ \bibinfo
  {pages} {237001} (\bibinfo {year} {2015})}\BibitemShut {NoStop}%
\bibitem [{\citenamefont {{Kim}}\ \emph {et~al.}()\citenamefont {{Kim}},
  \citenamefont {{Jip Park}},\ and\ \citenamefont {{Gilbert}}}]{gilbert2016}%
  \BibitemOpen
  \bibfield  {author} {\bibinfo {author} {\bibfnamefont {Y.}~\bibnamefont
  {{Kim}}}, \bibinfo {author} {\bibfnamefont {M.}~\bibnamefont {{Jip Park}}}, \
  and\ \bibinfo {author} {\bibfnamefont {M.~J.}\ \bibnamefont {{Gilbert}}},\
  }\href@noop {} {\ }\Eprint {http://arxiv.org/abs/arXiv:1604.01040}
  {arXiv:1604.01040} \BibitemShut {NoStop}%
\bibitem [{\citenamefont {Soto-Garrido}\ and\ \citenamefont
  {Fradkin}(2014)}]{rodrigo2015}%
  \BibitemOpen
  \bibfield  {author} {\bibinfo {author} {\bibfnamefont {R.}~\bibnamefont
  {Soto-Garrido}}\ and\ \bibinfo {author} {\bibfnamefont {E.}~\bibnamefont
  {Fradkin}},\ }\href {\doibase 10.1103/PhysRevB.89.165126} {\bibfield
  {journal} {\bibinfo  {journal} {Phys. Rev. B}\ }\textbf {\bibinfo {volume}
  {89}},\ \bibinfo {pages} {165126} (\bibinfo {year} {2014})}\BibitemShut
  {NoStop}%
\bibitem [{\citenamefont {{You}}\ and\ \citenamefont {{You}}(2016)}]{you2}%
  \BibitemOpen
  \bibfield  {author} {\bibinfo {author} {\bibfnamefont {Y.}~\bibnamefont
  {{You}}}\ and\ \bibinfo {author} {\bibfnamefont {Y.-Z.}\ \bibnamefont
  {{You}}},\ }\href@noop {} {\bibfield  {journal} {\bibinfo  {journal} {ArXiv
  e-prints}\ } (\bibinfo {year} {2016})},\ \Eprint
  {http://arxiv.org/abs/1601.00657} {arXiv:1601.00657 [cond-mat.str-el]}
  \BibitemShut {NoStop}%
\bibitem [{\citenamefont {Fradkin}\ \emph {et~al.}(2015)\citenamefont
  {Fradkin}, \citenamefont {Kivelson},\ and\ \citenamefont
  {Tranquada}}]{fradkin}%
  \BibitemOpen
  \bibfield  {author} {\bibinfo {author} {\bibfnamefont {E.}~\bibnamefont
  {Fradkin}}, \bibinfo {author} {\bibfnamefont {S.~A.}\ \bibnamefont
  {Kivelson}}, \ and\ \bibinfo {author} {\bibfnamefont {J.~M.}\ \bibnamefont
  {Tranquada}},\ }\href {\doibase 10.1103/RevModPhys.87.457} {\bibfield
  {journal} {\bibinfo  {journal} {Rev. Mod. Phys.}\ }\textbf {\bibinfo {volume}
  {87}},\ \bibinfo {pages} {457} (\bibinfo {year} {2015})}\BibitemShut
  {NoStop}%
\bibitem [{\citenamefont {Lee}(2014)}]{palee}%
  \BibitemOpen
  \bibfield  {author} {\bibinfo {author} {\bibfnamefont {P.~A.}\ \bibnamefont
  {Lee}},\ }\href {\doibase 10.1103/PhysRevX.4.031017} {\bibfield  {journal}
  {\bibinfo  {journal} {Phys. Rev. X}\ }\textbf {\bibinfo {volume} {4}},\
  \bibinfo {pages} {031017} (\bibinfo {year} {2014})}\BibitemShut {NoStop}%
\bibitem [{\citenamefont {Agterberg}\ \emph {et~al.}(2015)\citenamefont
  {Agterberg}, \citenamefont {Melchert},\ and\ \citenamefont
  {Kashyap}}]{agterberg}%
  \BibitemOpen
  \bibfield  {author} {\bibinfo {author} {\bibfnamefont {D.~F.}\ \bibnamefont
  {Agterberg}}, \bibinfo {author} {\bibfnamefont {D.~S.}\ \bibnamefont
  {Melchert}}, \ and\ \bibinfo {author} {\bibfnamefont {M.~K.}\ \bibnamefont
  {Kashyap}},\ }\href {\doibase 10.1103/PhysRevB.91.054502} {\bibfield
  {journal} {\bibinfo  {journal} {Phys. Rev. B}\ }\textbf {\bibinfo {volume}
  {91}},\ \bibinfo {pages} {054502} (\bibinfo {year} {2015})}\BibitemShut
  {NoStop}%
\bibitem [{\citenamefont {Wang}\ \emph
  {et~al.}(2015{\natexlab{a}})\citenamefont {Wang}, \citenamefont {Agterberg},\
  and\ \citenamefont {Chubukov}}]{pdw1}%
  \BibitemOpen
  \bibfield  {author} {\bibinfo {author} {\bibfnamefont {Y.}~\bibnamefont
  {Wang}}, \bibinfo {author} {\bibfnamefont {D.~F.}\ \bibnamefont {Agterberg}},
  \ and\ \bibinfo {author} {\bibfnamefont {A.}~\bibnamefont {Chubukov}},\
  }\href {\doibase 10.1103/PhysRevB.91.115103} {\bibfield  {journal} {\bibinfo
  {journal} {Phys. Rev. B}\ }\textbf {\bibinfo {volume} {91}},\ \bibinfo
  {pages} {115103} (\bibinfo {year} {2015}{\natexlab{a}})}\BibitemShut
  {NoStop}%
\bibitem [{\citenamefont {Wang}\ \emph
  {et~al.}(2015{\natexlab{b}})\citenamefont {Wang}, \citenamefont {Agterberg},\
  and\ \citenamefont {Chubukov}}]{pdw2}%
  \BibitemOpen
  \bibfield  {author} {\bibinfo {author} {\bibfnamefont {Y.}~\bibnamefont
  {Wang}}, \bibinfo {author} {\bibfnamefont {D.~F.}\ \bibnamefont {Agterberg}},
  \ and\ \bibinfo {author} {\bibfnamefont {A.}~\bibnamefont {Chubukov}},\
  }\href {\doibase 10.1103/PhysRevLett.114.197001} {\bibfield  {journal}
  {\bibinfo  {journal} {Phys. Rev. Lett.}\ }\textbf {\bibinfo {volume} {114}},\
  \bibinfo {pages} {197001} (\bibinfo {year} {2015}{\natexlab{b}})}\BibitemShut
  {NoStop}%
\bibitem [{\citenamefont {P\'epin}\ \emph {et~al.}(2014)\citenamefont
  {P\'epin}, \citenamefont {de~Carvalho}, \citenamefont {Kloss},\ and\
  \citenamefont {Montiel}}]{pdw_pepin}%
  \BibitemOpen
  \bibfield  {author} {\bibinfo {author} {\bibfnamefont {C.}~\bibnamefont
  {P\'epin}}, \bibinfo {author} {\bibfnamefont {V.~S.}\ \bibnamefont
  {de~Carvalho}}, \bibinfo {author} {\bibfnamefont {T.}~\bibnamefont {Kloss}},
  \ and\ \bibinfo {author} {\bibfnamefont {X.}~\bibnamefont {Montiel}},\ }\href
  {\doibase 10.1103/PhysRevB.90.195207} {\bibfield  {journal} {\bibinfo
  {journal} {Phys. Rev. B}\ }\textbf {\bibinfo {volume} {90}},\ \bibinfo
  {pages} {195207} (\bibinfo {year} {2014})}\BibitemShut {NoStop}%
\bibitem [{\citenamefont {{Hamidian}}\ \emph {et~al.}()\citenamefont
  {{Hamidian}}, \citenamefont {{Edkins}}, \citenamefont {{Joo}}, \citenamefont
  {{Kostin}}, \citenamefont {{Eisaki}}, \citenamefont {{Uchida}}, \citenamefont
  {{Lawler}}, \citenamefont {{Kim}}, \citenamefont {{Mackenzie}}, \citenamefont
  {{Fujita}}, \citenamefont {{Lee}},\ and\ \citenamefont {{S{\'e}amus
  Davis}}}]{davis_last}%
  \BibitemOpen
  \bibfield  {author} {\bibinfo {author} {\bibfnamefont {M.~H.}\ \bibnamefont
  {{Hamidian}}}, \bibinfo {author} {\bibfnamefont {S.~D.}\ \bibnamefont
  {{Edkins}}}, \bibinfo {author} {\bibfnamefont {S.~H.}\ \bibnamefont {{Joo}}},
  \bibinfo {author} {\bibfnamefont {A.}~\bibnamefont {{Kostin}}}, \bibinfo
  {author} {\bibfnamefont {H.}~\bibnamefont {{Eisaki}}}, \bibinfo {author}
  {\bibfnamefont {S.}~\bibnamefont {{Uchida}}}, \bibinfo {author}
  {\bibfnamefont {M.~J.}\ \bibnamefont {{Lawler}}}, \bibinfo {author}
  {\bibfnamefont {E.-A.}\ \bibnamefont {{Kim}}}, \bibinfo {author}
  {\bibfnamefont {A.~P.}\ \bibnamefont {{Mackenzie}}}, \bibinfo {author}
  {\bibfnamefont {K.}~\bibnamefont {{Fujita}}}, \bibinfo {author}
  {\bibfnamefont {J.}~\bibnamefont {{Lee}}}, \ and\ \bibinfo {author}
  {\bibfnamefont {J.~C.}\ \bibnamefont {{S{\'e}amus Davis}}},\ }\href@noop {}
  {\ }\Eprint {http://arxiv.org/abs/arXiv:1511.08124} {arXiv:1511.08124}
  \BibitemShut {NoStop}%
\bibitem [{\citenamefont {Burkov}\ \emph {et~al.}(2011)\citenamefont {Burkov},
  \citenamefont {Hook},\ and\ \citenamefont {Balents}}]{BHB2011}%
  \BibitemOpen
  \bibfield  {author} {\bibinfo {author} {\bibfnamefont {A.~A.}\ \bibnamefont
  {Burkov}}, \bibinfo {author} {\bibfnamefont {M.~D.}\ \bibnamefont {Hook}}, \
  and\ \bibinfo {author} {\bibfnamefont {L.}~\bibnamefont {Balents}},\ }\href
  {\doibase 10.1103/PhysRevB.84.235126} {\bibfield  {journal} {\bibinfo
  {journal} {Phys. Rev. B}\ }\textbf {\bibinfo {volume} {84}},\ \bibinfo
  {pages} {235126} (\bibinfo {year} {2011})}\BibitemShut {NoStop}%
\bibitem [{\citenamefont {Maciejko}\ and\ \citenamefont
  {Nandkishore}(2014)}]{nandkishore2014}%
  \BibitemOpen
  \bibfield  {author} {\bibinfo {author} {\bibfnamefont {J.}~\bibnamefont
  {Maciejko}}\ and\ \bibinfo {author} {\bibfnamefont {R.}~\bibnamefont
  {Nandkishore}},\ }\href {\doibase 10.1103/PhysRevB.90.035126} {\bibfield
  {journal} {\bibinfo  {journal} {Phys. Rev. B}\ }\textbf {\bibinfo {volume}
  {90}},\ \bibinfo {pages} {035126} (\bibinfo {year} {2014})}\BibitemShut
  {NoStop}%
\bibitem [{\citenamefont {Sur}\ and\ \citenamefont {Lee}(2014)}]{Sur_lee2014}%
  \BibitemOpen
  \bibfield  {author} {\bibinfo {author} {\bibfnamefont {S.}~\bibnamefont
  {Sur}}\ and\ \bibinfo {author} {\bibfnamefont {S.-S.}\ \bibnamefont {Lee}},\
  }\href {\doibase 10.1103/PhysRevB.90.045121} {\bibfield  {journal} {\bibinfo
  {journal} {Phys. Rev. B}\ }\textbf {\bibinfo {volume} {90}},\ \bibinfo
  {pages} {045121} (\bibinfo {year} {2014})}\BibitemShut {NoStop}%
\bibitem [{\citenamefont {Chan}\ \emph {et~al.}()\citenamefont {Chan},
  \citenamefont {Chiu}, \citenamefont {Chou},\ and\ \citenamefont
  {Schnyder}}]{Chan:2015aa}%
  \BibitemOpen
  \bibfield  {author} {\bibinfo {author} {\bibfnamefont {Y.-H.}\ \bibnamefont
  {Chan}}, \bibinfo {author} {\bibfnamefont {C.-K.}\ \bibnamefont {Chiu}},
  \bibinfo {author} {\bibfnamefont {M.~Y.}\ \bibnamefont {Chou}}, \ and\
  \bibinfo {author} {\bibfnamefont {A.~P.}\ \bibnamefont {Schnyder}},\
  }\href@noop {} {\ }\Eprint {http://arxiv.org/abs/arXiv:1510.02759}
  {arXiv:1510.02759} \BibitemShut {NoStop}%
\bibitem [{\citenamefont {Bian}\ \emph {et~al.}(2016)\citenamefont {Bian},
  \citenamefont {Chang}, \citenamefont {Zheng}, \citenamefont {Velury},
  \citenamefont {Xu}, \citenamefont {Neupert}, \citenamefont {Chiu},
  \citenamefont {Huang}, \citenamefont {Sanchez}, \citenamefont {Belopolski},
  \citenamefont {Alidoust}, \citenamefont {Chen}, \citenamefont {Chang},
  \citenamefont {Bansil}, \citenamefont {Jeng}, \citenamefont {Lin},\ and\
  \citenamefont {Hasan}}]{bian_2016}%
  \BibitemOpen
  \bibfield  {author} {\bibinfo {author} {\bibfnamefont {G.}~\bibnamefont
  {Bian}}, \bibinfo {author} {\bibfnamefont {T.-R.}\ \bibnamefont {Chang}},
  \bibinfo {author} {\bibfnamefont {H.}~\bibnamefont {Zheng}}, \bibinfo
  {author} {\bibfnamefont {S.}~\bibnamefont {Velury}}, \bibinfo {author}
  {\bibfnamefont {S.-Y.}\ \bibnamefont {Xu}}, \bibinfo {author} {\bibfnamefont
  {T.}~\bibnamefont {Neupert}}, \bibinfo {author} {\bibfnamefont {C.-K.}\
  \bibnamefont {Chiu}}, \bibinfo {author} {\bibfnamefont {S.-M.}\ \bibnamefont
  {Huang}}, \bibinfo {author} {\bibfnamefont {D.~S.}\ \bibnamefont {Sanchez}},
  \bibinfo {author} {\bibfnamefont {I.}~\bibnamefont {Belopolski}}, \bibinfo
  {author} {\bibfnamefont {N.}~\bibnamefont {Alidoust}}, \bibinfo {author}
  {\bibfnamefont {P.-J.}\ \bibnamefont {Chen}}, \bibinfo {author}
  {\bibfnamefont {G.}~\bibnamefont {Chang}}, \bibinfo {author} {\bibfnamefont
  {A.}~\bibnamefont {Bansil}}, \bibinfo {author} {\bibfnamefont {H.-T.}\
  \bibnamefont {Jeng}}, \bibinfo {author} {\bibfnamefont {H.}~\bibnamefont
  {Lin}}, \ and\ \bibinfo {author} {\bibfnamefont {M.~Z.}\ \bibnamefont
  {Hasan}},\ }\href {\doibase 10.1103/PhysRevB.93.121113} {\bibfield  {journal}
  {\bibinfo  {journal} {Phys. Rev. B}\ }\textbf {\bibinfo {volume} {93}},\
  \bibinfo {pages} {121113} (\bibinfo {year} {2016})}\BibitemShut {NoStop}%
\bibitem [{\citenamefont {Matsuura}\ \emph {et~al.}(2013)\citenamefont
  {Matsuura}, \citenamefont {Chang}, \citenamefont {Schnyder},\ and\
  \citenamefont {Ryu}}]{Matsuura2013}%
  \BibitemOpen
  \bibfield  {author} {\bibinfo {author} {\bibfnamefont {S.}~\bibnamefont
  {Matsuura}}, \bibinfo {author} {\bibfnamefont {P.-Y.}\ \bibnamefont {Chang}},
  \bibinfo {author} {\bibfnamefont {A.~P.}\ \bibnamefont {Schnyder}}, \ and\
  \bibinfo {author} {\bibfnamefont {S.}~\bibnamefont {Ryu}},\ }\href
  {http://stacks.iop.org/1367-2630/15/i=6/a=065001} {\bibfield  {journal}
  {\bibinfo  {journal} {New J. Phys.}\ }\textbf {\bibinfo {volume} {15}},\
  \bibinfo {pages} {065001} (\bibinfo {year} {2013})}\BibitemShut {NoStop}%
\bibitem [{\citenamefont {Ramamurthy}\ and\ \citenamefont
  {Hughes}()}]{Ramamurthy:2015aa}%
  \BibitemOpen
  \bibfield  {author} {\bibinfo {author} {\bibfnamefont {S.~T.}\ \bibnamefont
  {Ramamurthy}}\ and\ \bibinfo {author} {\bibfnamefont {T.~L.}\ \bibnamefont
  {Hughes}},\ }\href@noop {} {\ }\Eprint
  {http://arxiv.org/abs/arXiv:1508.01205} {arXiv:1508.01205} \BibitemShut
  {NoStop}%
\bibitem [{\citenamefont {Zyuzin}\ \emph {et~al.}(2012)\citenamefont {Zyuzin},
  \citenamefont {Wu},\ and\ \citenamefont {Burkov}}]{burkov2012}%
  \BibitemOpen
  \bibfield  {author} {\bibinfo {author} {\bibfnamefont {A.~A.}\ \bibnamefont
  {Zyuzin}}, \bibinfo {author} {\bibfnamefont {S.}~\bibnamefont {Wu}}, \ and\
  \bibinfo {author} {\bibfnamefont {A.~A.}\ \bibnamefont {Burkov}},\ }\href
  {\doibase 10.1103/PhysRevB.85.165110} {\bibfield  {journal} {\bibinfo
  {journal} {Phys. Rev. B}\ }\textbf {\bibinfo {volume} {85}},\ \bibinfo
  {pages} {165110} (\bibinfo {year} {2012})}\BibitemShut {NoStop}%
    \bibitem{xuzhangzhang2015}Yong Xu, Fan Zhang, and Chuanwei Zhang, Phys. Rev. Lett. \textbf{115}, 265304 (2015).
\bibitem{bhyanpressure} R. D. dos Reis, S. C. Wu, Y. Sun, M. O. Ajeesh, C. Shekhar, M. Schmidt, C. Felser, B. Yan, and M. Nicklas
Phys. Rev. B \textbf{93}, 205102 (2016).
\bibitem [{\citenamefont {Vazifeh}\ and\ \citenamefont
  {Franz}(2013)}]{franz2013}%
  \BibitemOpen
\bibfield  {journal} {  }\bibfield  {author} {\bibinfo {author} {\bibfnamefont
  {M.~M.}\ \bibnamefont {Vazifeh}}\ and\ \bibinfo {author} {\bibfnamefont
  {M.}~\bibnamefont {Franz}},\ }\href {\doibase 10.1103/PhysRevLett.111.027201}
  {\bibfield  {journal} {\bibinfo  {journal} {Phys. Rev. Lett.}\ }\textbf
  {\bibinfo {volume} {111}},\ \bibinfo {pages} {027201} (\bibinfo {year}
  {2013})}\BibitemShut {NoStop}%
\bibitem [{\citenamefont {Wang}\ and\ \citenamefont
  {Zhang}(2013)}]{Wang2013axion}%
  \BibitemOpen
  \bibfield  {author} {\bibinfo {author} {\bibfnamefont {Z.}~\bibnamefont
  {Wang}}\ and\ \bibinfo {author} {\bibfnamefont {S.-C.}\ \bibnamefont
  {Zhang}},\ }\href {\doibase 10.1103/PhysRevB.87.161107} {\bibfield  {journal}
  {\bibinfo  {journal} {Phys. Rev. B}\ }\textbf {\bibinfo {volume} {87}},\
  \bibinfo {pages} {161107} (\bibinfo {year} {2013})}\BibitemShut {NoStop}%
\bibitem [{\citenamefont {Roy}\ and\ \citenamefont {Sau}(2015)}]{roy_sau2015}%
  \BibitemOpen
  \bibfield  {author} {\bibinfo {author} {\bibfnamefont {B.}~\bibnamefont
  {Roy}}\ and\ \bibinfo {author} {\bibfnamefont {J.~D.}\ \bibnamefont {Sau}},\
  }\href {\doibase 10.1103/PhysRevB.92.125141} {\bibfield  {journal} {\bibinfo
  {journal} {Phys. Rev. B}\ }\textbf {\bibinfo {volume} {92}},\ \bibinfo
  {pages} {125141} (\bibinfo {year} {2015})}\BibitemShut {NoStop}%
\bibitem [{\citenamefont {{Redell}}\ \emph {et~al.}()\citenamefont {{Redell}},
  \citenamefont {{Mukherjee}},\ and\ \citenamefont {{Lee}}}]{wclee2016}%
  \BibitemOpen
  \bibfield  {author} {\bibinfo {author} {\bibfnamefont {M.~D.}\ \bibnamefont
  {{Redell}}}, \bibinfo {author} {\bibfnamefont {S.}~\bibnamefont
  {{Mukherjee}}}, \ and\ \bibinfo {author} {\bibfnamefont {W.-C.}\ \bibnamefont
  {{Lee}}},\ }\href@noop {} {\ }\Eprint {http://arxiv.org/abs/arXiv:1603.06810}
  {arXiv:1603.06810} \BibitemShut {NoStop}%
 \bibitem{you_cho_hughes2016} Y.\ You, G.\ Y.\ Cho, and T.\ L.\ Hughes, arXiv:1605.02734.
\bibitem [{foo()}]{footnote}%
  \BibitemOpen
  \href@noop {} {\bibinfo  {journal} {From these projective form factors, one
  can also obtain the nodal structures if multiple orders coexist. For example,
  the coexistence of SDW$_x$ and SDW$_y$ can either have line nodes or point
  nodes, depending on the relative phase of the order parameters}\
  }\BibitemShut {NoStop}%
\bibitem [{\citenamefont {Qi}\ \emph {et~al.}(2008)\citenamefont {Qi},
  \citenamefont {Hughes},\ and\ \citenamefont {Zhang}}]{Qi2008}%
  \BibitemOpen
\bibfield  {journal} {  }\bibfield  {author} {\bibinfo {author} {\bibfnamefont
  {X.-L.}\ \bibnamefont {Qi}}, \bibinfo {author} {\bibfnamefont {T.~L.}\
  \bibnamefont {Hughes}}, \ and\ \bibinfo {author} {\bibfnamefont {S.-C.}\
  \bibnamefont {Zhang}},\ }\href {\doibase 10.1103/PhysRevB.78.195424}
  {\bibfield  {journal} {\bibinfo  {journal} {Phys. Rev. B}\ }\textbf {\bibinfo
  {volume} {78}},\ \bibinfo {pages} {195424} (\bibinfo {year}
  {2008})}\BibitemShut {NoStop}%
\bibitem [{\citenamefont {{Wang}}\ \emph
  {et~al.}(2016{\natexlab{b}})\citenamefont {{Wang}}, \citenamefont {{Cho}},
  \citenamefont {{Hughes}},\ and\ \citenamefont {{Fradkin}}}]{wang_new}%
  \BibitemOpen
  \bibfield  {author} {\bibinfo {author} {\bibfnamefont {Y.}~\bibnamefont
  {{Wang}}}, \bibinfo {author} {\bibfnamefont {G.~Y.}\ \bibnamefont {{Cho}}},
  \bibinfo {author} {\bibfnamefont {T.~L.}\ \bibnamefont {{Hughes}}}, \ and\
  \bibinfo {author} {\bibfnamefont {E.}~\bibnamefont {{Fradkin}}},\ }\href@noop
  {} {\bibfield  {journal} {\bibinfo  {journal} {ArXiv e-prints}\ } (\bibinfo
  {year} {2016}{\natexlab{b}})},\ \Eprint {http://arxiv.org/abs/1602.02778}
  {arXiv:1602.02778 [cond-mat.str-el]} \BibitemShut {NoStop}%
\bibitem [{\citenamefont {Murakami}\ and\ \citenamefont
  {Nagaosa}(2003)}]{nagaosa}%
  \BibitemOpen
  \bibfield  {author} {\bibinfo {author} {\bibfnamefont {S.}~\bibnamefont
  {Murakami}}\ and\ \bibinfo {author} {\bibfnamefont {N.}~\bibnamefont
  {Nagaosa}},\ }\href {\doibase 10.1103/PhysRevLett.90.057002} {\bibfield
  {journal} {\bibinfo  {journal} {Phys. Rev. Lett.}\ }\textbf {\bibinfo
  {volume} {90}},\ \bibinfo {pages} {057002} (\bibinfo {year}
  {2003})}\BibitemShut {NoStop}%
\bibitem [{\citenamefont {Bernevig}\ and\ \citenamefont
  {Hughes}(2013)}]{bernevig_book}%
  \BibitemOpen
  \bibfield  {author} {\bibinfo {author} {\bibfnamefont {B.~A.}\ \bibnamefont
  {Bernevig}}\ and\ \bibinfo {author} {\bibfnamefont {T.~L.}\ \bibnamefont
  {Hughes}},\ }\href@noop {} {\emph {\bibinfo {title} {Topological Insulators
  and Topological Superconductors}}}\ (\bibinfo  {publisher} {Princeton
  University Press},\ \bibinfo {year} {2013})\BibitemShut {NoStop}%
\end{thebibliography}
\end{document}